\documentstyle[aps,epsfig,eqsecnum]{revtex}
\firstfigfalse
 \draft
\tighten
\begin{document}
\title{Paraxial propagation of a quantum charge\\ in a random magnetic field}
\author{A. Shelankov$^{*}$}
\address{Department of Theoretical Physics, Ume{\aa} University, 901 87
Ume{\aa}, Sweden}

\date{version2: 28 April 2000}

\maketitle

\begin{abstract}
The paraxial (parabolic) theory of a near forward scattering of a
quantum charged particle by a static magnetic field is presented.
From the paraxial solution to the Aharonov-Bohm scattering problem the
transverse transfered momentum (the Lorentz force) is found.  Multiple
scattering is considered for two models: (i) Gaussian $\delta
$-correlated random magnetic field; (ii) a random array of the
Aharonov-Bohm magnetic flux line.  The paraxial gauge-invariant
two-particle Green function averaged with respect to the random
magnetic field is found by an exact evaluation of the Feynman
integral.  It is shown that in spite of the anomalous character of the
forward scattering, the transport properties can be described by the
Boltzmann equation. The Landau quantization in the gauge field of the
Aharonov-Bohm lines is discussed.  
\end{abstract}


\section{Introduction}
\label{Intro}

The paper addresses the problem of quantum transport of a charge in an
inhomogeneous static random magnetic field.  In recent years, this or
related problems have been met in a number of contexts in physics of
2-dimensional systems.  For instance, in the composite fermion model
of the Fractional Quantum Hall Effect, a (fictitious) random magnetic
field is the environment which controls dynamics of effective charge
carriers \cite{QHE}.  One meets the fluctuating gauge fields in some
models of high temperature superconductors \cite{NagLee}, where the
gauge field is the tool to impose the constrain of no double occupancy
in the $t-J$ model \cite{highTc}.  Besides, stochastically
inhomogeneous magnetic field can be experimentally created by various
ways.  For example, the field is irregular near the surface of a
superconductor in an external magnetic field if the Abrikosov flux
lattice is disordered; depending on the experimental conditions, the
magnetic field inhomogeneities may be weak and smooth, or the field
may be concentrated in an irregular array of flux tubes. Various
aspects of transport in the magnetic field of the Abrikosov vortices,
the weak localization in particular, have been studied in Refs.
\cite{RamShe87,Gei89,BenKliPlo90,GeiDubKha90,GeiFalDub92,GeiBenGri92,GeiBenGri94}.
In recent years, the random magnetic field problem has been an active
subject area Refs.
\cite{AltIof92,KhvMes93,AroMirWol94,AroAltMir95-1,AroAltMir95-2,NieHed95,%
MirPolWol98,EveMirPol99,TarEfe00}.

The formulation of the problem is as follows.  A particle with the
electric charge $e$ and the mass $m$ moves on the $x-y$ plane subject
to a vector potential potential $\bbox{A}(A_{x},A_{y})$ generated by a
magnetic field $ b(x,y)=\left(\bbox{{\rm rot\,} A} \right)_{z}$.  Two
random fields models are considered in the paper. In the first one,
the magnetic field $b(\bbox{r})$ is a random Gaussian variable with
zero average, $\langle b \rangle =0 $, specified by the correlator
\begin{equation}
\langle b(\bbox{r})b(\bbox{r}') \rangle  = 
\left( {\Phi _{0}\over{2 \pi }}  \right) ^{2}
{1\over{{\cal L}^{2}}} \delta (\bbox{r}-\bbox{r}') \ , 
\label{vhb}
\end{equation}
where $\Phi_{0}= {hc\over e}$ is the flux quantum; the strength of the
random magnetic field is characterized via the length ${\cal L}$ the
meaning of which is that the magnetic flux through the area ${\cal
L}^{2}$ is typically of order of $\Phi_{0}$.  The random field is
assumed to be weak in the sense that $ {\cal L}$ much exceeds the wave
length $\lambdabar \equiv \hbar /p $, $p$ being the particle momentum.
In another model \cite{EmpBas94,DesFurOuv95,DesOuvTex97} which is
motivated by fractional statistics theories, the gauge potential is
created by a random array of the Aharonov-Bohm flux lines.  A system
of the Abrikosov vortices (e.g. in the gate of a MOSFET transistor
\cite{RamShe87,Gei89,BenKliPlo90}) may serve as an experimental
realization of the Aharonov-Bohm array if the particle wave length
much exceeds the vortex (magnetic) size.

In the random magnetic field case, the traditional approach of the
theory of disordered systems \cite{AbrGorDzy63} meets difficulties on
the very first steps. Indeed, the simplest object that is the single
particle Green's function $G(1,2)=\langle \psi
(\bbox{r}_{1})\psi^{*}(\bbox{r}_{2}) \rangle $, is not gauge invariant
and the physical meaning of its averaging with respect to the vector
potential $\bbox{A}$ generated by the random field is not clear. One
may define a gauge invariant combination $\tilde{G}(1,2) = \langle
\psi (\bbox{r}_{1})\psi^{*}(\bbox{r}_{2}) \rangle \exp[i {e \hbar
\over c} \int_{{\cal C}_{12}}d \bbox{l\cdot A}] $ where the path
${\cal C}_{12}$ connects the points $\bbox{r}_{1}$ and $\bbox{r}_{2}$.
Albeit gauge independent, $\tilde{G}(1,2)$ essentially depends on the
choice of the path ${\cal C}_{12}$.  With the point $\bbox{r}_{1,2}$
connected by the straight line, the field averaged $\tilde{G}(1,2)$
has been found in Ref.\cite{AltIof92}.

Another problem is the diverging scattering rate ${1\over \tau }$.
For small scattering angles $\phi $, the differential cross-section
behaves like ${1\over \phi^{2}}$ so that the scattering total
cross-section is infinite.  In other words, the life time of a state
with the definite momentum is zero.  The conventional diagram
technique \cite{AbrGorDzy63} where ${\hbar \over \tau }$ is assumed to
be small compared with the kinetic energy, becomes questionable.  On
the other hand, it is known \cite{KimFurWen94} that in gauge-invariant
response functions the self energy enters in combination with the
vertex corrections and the divergence cancels out.  In
Ref.\cite{AroAltMir95-1}, it has been attempted to introduce a
physically sensible gauge-invariant ``single-particle time'' $\tau $
as a parameter entering the Landau level broadening.

The main purpose of this paper is to develop a scheme which allows one
to study the most singular part of interaction with random magnetic
field, that is the near forward scattering.

To pinpoint the physics behind the theoretical difficulties, consider
first propagation of a plane wave.  It is common in wave mechanics to
analyze propagation in terms of wave fronts, {\it i.e.}  the surfaces
(lines in 2D) of constant phase.  The property of the wave front line
is that the probability current is locally perpendicular to the line.
In the magnetic field, the phase of the wave function is ill-defined
because of the gauge freedom.  Nevertheless, one can construct a gauge
invariant quantity $\chi $ defined on a line, which in a limited sense
plays the role of the phase: Given the wave function $\psi(\bbox{r})$
and the vector potential $\bbox{A}$, the phase $\chi (\bbox{s})$ for
the points $\bbox{s}$ on a line $S$ is defined through its
differential as
\begin{equation}
d \chi = {m\over \hbar |\psi|^{2}} \bbox{j\cdot}d \bbox{s}
\label{3hb}
\end{equation}
where $\bbox{j} = {1 \over m} \Re \left(\psi^{*} \left({\hbar\over
i}\bbox{\nabla} - {e\over c} \bbox{A} \right) \psi \right) $ is the
probability current density.  Provided the line $S$ does not have
self-intersections, $\chi $ is an unique function of $\bbox{s}$.

If $\chi$ is a constant, {\it i.e.} $\bbox{j\cdot}d \bbox{s}=0$ and
$\bbox{j}\perp d \bbox{s}$, the local current and the normal to $S$
are parallel, so that $S$ is a wave front.  If $\chi $ is a slowly
varying function, $\lambdabar \left(\chi (\bbox{s})- \chi (\bbox{0})
\right) $ gives the local distance from $S$ to the wave front passing
through the point $\bbox{s}=0$.

\begin{figure}[h] 
\centerline{\epsfig{file=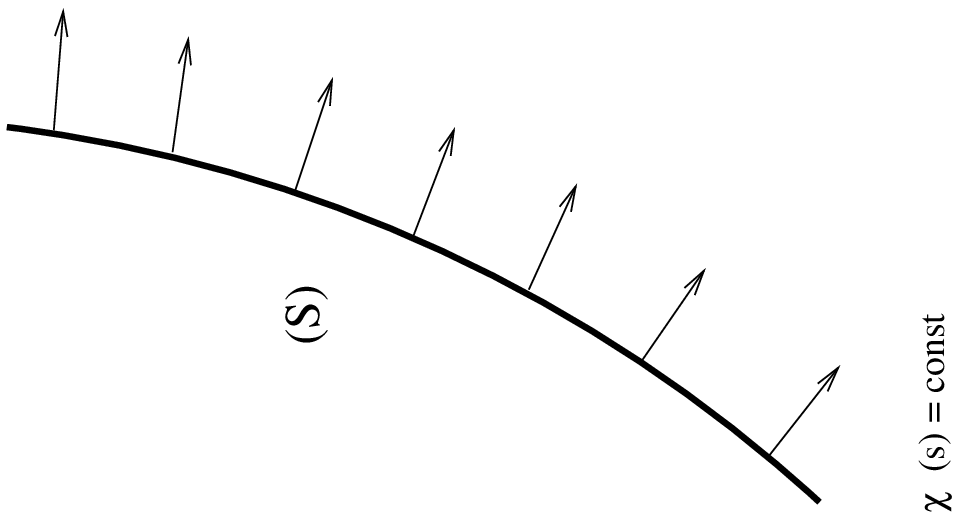,width=0.20\textwidth,%
angle=-90}
\hspace*{0.2\textwidth}
\epsfig{file=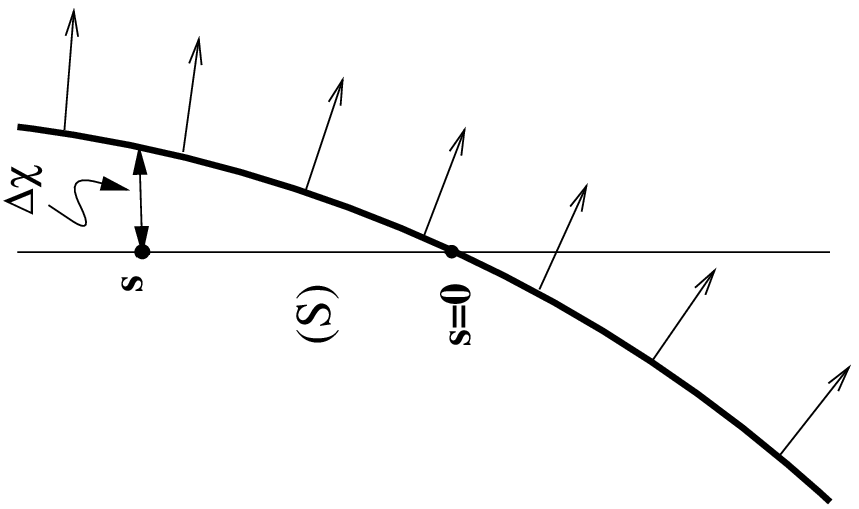,width=0.20\textwidth,%
angle=-90}
}
\centerline{(a)\hspace*{0.5\textwidth}(b)}

\caption{The wave front of a wave on a plane. (a) The wave front $S$
is a line at each point of which the probability current is directed
perpendicular to the line. The gauge invariant phase $\chi $ is a
constant along a wave front.  On a general line $S$, chosen as a
straight line in (b), the phase a function of the coordinate along the
line $\bbox{s}$. Its variation is found from Eq.(\ref{3hb}).  The
physical meaning of $\chi (\bbox{s})$ is that the phase difference
$\Delta \chi (\bbox{s})\equiv \left(\chi (\bbox{s})- \chi (\bbox{0})
\right)$ multiplied by the wave length $\lambdabar $ shows the local
distance $\lambdabar \Delta \chi $ from the line $S$ to the wave
front.} 
\label{front}
\end{figure}

Consider now how the random magnetic field affects the wave front upon
its propagation.  Take a state $\psi $, for which the line $x=0$ is a
wave front corresponding to the propagation in the positive
$x$-direction ($j_{x}$ >0).  To satisfy the requirement $ d \chi =0$,
the wave function is
\[
\psi (x=0, y)= \exp\left[i {e \hbar \over c}
\int\limits_{-\infty }^{y} dy \; A_{y}(x=0,y)\right] \; ,
\]
along the wave front $x=0$ (choice of lower limit of integration is
not important).

\begin{figure}[h] 
\centerline{\epsfig{file=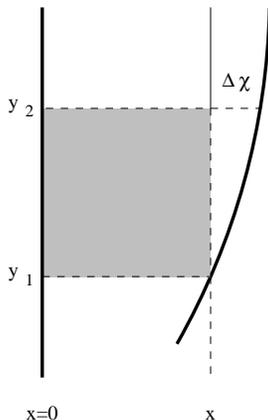,height=0.15\textheight,angle=-90}}
\caption{Propagation of the wave front magnetic field. 
As discussed in the text, one constructs a state for which the
straight line at $x=0$ is a wave front.  For a finite $x$, the
profile of the front passing via the point $x,y_{1}$ is controlled by the
phase difference $\Delta \chi = \chi (y_{2};x)- \chi (y_{1};z)$. The
difference is proportional to the magnetic flux through
through the shaded area.
}
\label{magnFront}
\end{figure}

To find the profile of the wave front having advanced from $x=0$ to a
finite $x$, one can apply the usual eikonal-type approximation where
the field affects only the phase of the wave function through the
factor $\exp[{ie\over \hbar c}\int d\bbox{l\cdot A}]$, $ d \bbox{l}
\parallel \hat{x}$ being along the direction of propagation:
\[
\psi ( x, y) = \exp\left[{i\over{\hbar}}
\left(p x + 
{e\over c}\int\limits_{-\infty }^{y} dy \; A_{y}(0,y)
+{e\over 
c}\int\limits_{0}^{x}dx\;A_{x}(x,y) \right)
\right]\; .
\]
Integrating Eq.(\ref{3hb}), one finds the phase $\chi(y;x) $ as a
function of $y$ for fixed $x$. For the phase difference $\Delta \chi
(y_{1},y_{2};x)=\chi ( y_{2};x) - \chi(y_{1}; x)$, one gets after
simple calculations:
\begin{equation}
\Delta \chi (y_{1},y_{2};x)
 =
2\pi  { \Phi(y_{1},y_{2}; x)\over \Phi_{0}}
\label{4hb}
\end{equation}
where $\Phi(y_{1},y_{2}; x)$ is the magnetic flux through the area
enclosed by the path $(0, y_{1}) \rightarrow (0,y_{2}) \rightarrow (
x,y_{2})\rightarrow ( x,y_{1}) \rightarrow (0, y_{1})$ (see
fig. \ref{magnFront}).

A non-local character of the interaction with magnetic filed is
clearly seen from Eq.(\ref{4hb}): the phase difference is controlled
by the flux rather than the magnetic field in the vicinity of the
particle trajectories. The non-locality is obviously of the
Aharonov-Bohm type.

Averaging with respect of the random magnetic field Eq.(\ref{vhb}),
one gets the variation of the phase difference
\[
(\Delta \chi)^{2}(y_{1},y_{2};x) = {1\over 2} {|x| \cdot
|y_{1}-y_{2}|\over {\cal L}^{2}}\; . 
\]
Most notable feature here is that $(\Delta \chi)^{2}$ grows with the
separation $|y_{1}-y_{2}|$ (cf. Ref.\cite{AltIof92}).  For points
separated by the distance $\Delta y$, the random phase difference is
of order of $1 $ when the wave front advances to $\Delta x \sim {\cal
L}^{2}/ \Delta y$.  One sees that an infinite plane wave, for which
$\Delta y \rightarrow \infty $, looses its coherence immediately,
whatever small is the propagation distance $\Delta x$.  (See Section
\ref{random} for a more formal derivation.)  These qualitative
arguments explain the actual meaning of the zero life time and show
that it is not an artifact arising e.g. due to a violation of the
gauge invariance.

To handle the anomalously intensive forward scattering, one needs a
method suitable for a nonperturbative analysis of the small-angle
multiple scattering.  For this, the paraxial (parabolic) approximation
\cite{parax} to the Schr\"odinger equation is chosen in the paper.
The paraxial theory is applicable when the particle moves mainly in
the direction of an ``axis'' and the momentum transverse to the axis
remains always small.  The paraxial approximation to the wave equation
is most popular in optics where it gives a convenient description of
light beams propagating in optical systems, their diffraction,
focusing etc \cite{Arn76}.  Taking scattering and diffraction
broadening on equal footing, the paraxial approximation is more
generally applicable then the eikonal one \cite{LanLif3}.

To make the paper self-contained, the derivation of the paraxial
approximation is outlined in Sect.\ref{parax}.  The case of magnetic
field is considered in Sect.\ref{magn} where a scheme for description
of scattering by magnetic field is suggested.  The scheme is in a
sense gauge invariant, gauge freedom revealing itself only in the
overall phases.  As a limiting case, one recovers the well-known
eikonal approximation (see Sect. \ref{eikonal}).

To illustrate usage of the paraxial approximation, a simple problem of
scattering of a charged particle by the Aharonov-Bohm magnetic flux
line is considered in Sect.\ref{AB}.  This (or equivalent) problem is
of interest in a broad variety of contexts extending from the cosmic
string theory \cite{Vol98} to superfluids \cite{Ior65,Son75,Son97}
(the Iordanskii force) and superconductors \cite{Cle68} where the
scattering of excitations by quantized vortex lines controls the
vortex dynamics.  Although the exact solution to the scattering
problem has been known since the original paper of Aharonov and Bohm
\cite{AhaBoh59} (see also review \cite{OlaPop85} and references
therein), certain controversy in the analysis and the interpretation
of the solution still remains.  Different opinions exist in the
literature about the existence of the transverse force exerted by the
Aharonov-Bohm line or a superfluid vortex.  On the basis of the
left-right symmetry in the Aharonov-Bohm differential cross-section,
the authors of Refs. \cite{MarWil88} and Ref.\cite{ThoAoNiu97} have
come to the conclusion that the line does not exert any Lorentz-like
force (translated as the Iordanskii force in a superfluid).  Other
authors, \cite{Son97,EmpBas94,DesOuvTex97,She98,Ber99} predict a
finite force.  Due to its simplicity, the paraxial solution allows one
to perform a detailed analysis and resolve the controversy.

It is shown in In Sect.\ref{dens}, that the paraxial scattering theory
becomes manifestly gauge invariant if formulated in terms of by-linear
in $\psi $ and $\psi^{*}$ object that is the density matrix $\rho $.
The evolution of the density matrix is given by a gauge invariant
two-particle Green function.

As discussed in Sect.\ref{path}, the paraxial 2D stationary equation
with inhomogeneous magnetic field can be written as a non-stationary
1D Schr\"odinger equation for a particle in a time-dependent electric
field.  This mapping allows one to present stationary solutions to the
2D magnetic field problem as the Feynman path integral for the
effective 1D problem.
 
In Sect.\ref{random}, the paraxial theory is applied to the model of
$\delta$-correlated random magnetic field.  It turns out to be
possible to evaluate the Feynman path integral and by this to find a
(paraxially) exact expression for the two-particle Green's function
averaged with respect to the magnetic field fluctuations.  It is shown
that the density matrix evolution can be mapped to the Boltzmann
kinetic equation.

In Sect.\ref{AB-array}, another model is considered where the random
gauge filed is generated by a random array of Aharonov-Bohm fluxes.
The flux lines are randomly distributed in the plane, the flux of a
line, $\Phi$, is distributed with the probability $p(\Phi )$.  The
Aharonov-Bohm array may create an effective magnetic field $\tilde{B}$
if $p(\Phi )$ is asymmetric, $p(\Phi )\neq p(-\Phi )$.

In Sect.\ref{kin}, the Boltzmann equation for charge subject to a
Gaussian random magnetic field or field of AB-array is derived.  With
the help of the Boltzmann equation, the resistivity tensor is found.
Finally, in Sect.\ref{Landau} we discuss the density of states of the
levels due to the quantization of motion in the field $\tilde{B}$.

The results are summarized in  Section \ref{concl}

\section{Paraxial approximation}
\label{parax}

The paraxial approximation allows one to construct a family of
solutions to the Schr\"odinger equation which are close to to the
plane wave with a certain momentum $\bbox{p}_{0}$.  The wave function
$\Psi_{Sch}$, a solution to the stationary Schr\"odinger equation, is
presented as
\begin{equation}
\Psi_{Sch}(\bbox{r})= 
\Psi (\bbox{r})
e^{{i\over{\hbar}}\bbox{p}_{0}\bbox{\cdot r} }\ ,
\label{mca}
\end{equation}
where the envelope paraxial function $\Psi (\bbox{r})$ is supposed to
be slowly varying at the distances of order of the wave length
$\lambdabar = {\hbar \over{p_{0}}}$.

The Schr\"odinger equation reads
\begin{equation}
\left( E  (\hat{\bbox{p}})+ U \right)\Psi_{Sch}= E \Psi_{Sch}
\label{gca}
\end{equation}
$ E (\bbox{p})= {1\over{2m}}\bbox{p}^{2}$ and $U(\bbox{r})$ being the
kinetic and potential energy respectively.  Given $\bbox{p}_{0}$, the
family of solutions in Eq.(\ref{mca}) corresponds to the eigen-energy
$E= E(\bbox{p}_{0})$ and the velocity $\bbox{v}= {\partial E
(\bbox{p}_{0})\over{\partial \bbox{p}_{0} }}$.  Inserting
Eq.(\ref{mca}) into Eq.(\ref{gca}), one gets equation for $\Psi $,
\[
 \Big(\widetilde E\big(\hat{\bbox{p}}\big) - E(\bbox{p}_{0})
 + U\Big)\Psi = 0 \;\; , \;\;
  \widetilde E\big(\hat{\bbox{p}}\big)\equiv 
e^{- {i\over{\hbar}} \bbox{p}_{0}\bbox{\cdot r}}
 E  \big(\hat{\bbox{p}}\big)
e^{{i\over{\hbar}} \bbox{p}_{0}\bbox{\cdot r}}\; , 
\]
which is still exact.  The operator $\widetilde{E}(\hat{\bbox{p}})=E(
\bbox{p}_{0}+ {\hbar \over{i}} \bbox{\nabla })$ acting on the slowly
varying function $\Psi $ is approximated in the paraxial theory as
\[
\widetilde{E}(\hat{\bbox{p}})\approx E(\bbox{p}_{0})+
{\hbar\over{i}}\bbox{v} \bbox{\cdot \nabla} + {1\over{2m}}
\left({\hbar\over{i}} \bbox{\nabla }_{\perp}\right)^{2}\; ,
\]
here $\bbox{\nabla}_{\perp}$ denotes the gradient in the direction
perpendicular to $\bbox{v}$.  

The paraxial approximation to
the Schr\"odinger equation reads
\begin{equation}
\left( 
  i \hbar \bbox{v} \bbox{\cdot \nabla} + {\hbar ^{2}\over{2m}}
  \bbox{\nabla }_{\perp}^{2} - U \right)\Psi =0 \; .
\label{hca}
\end{equation}
(Condition of applicability  are discussed later).
The main feature of the paraxial approximation Eq.(\ref{hca}), is that
it is of first order differential equation relative to the coordinate in
the direction of the propagation $x= \bbox{r\cdot v}/v$.

Introducing formally  ``time'' $\tau = x/v$, Eq.(\ref{hca}) takes the form,
of the time dependent Schr\"odinger equation in a reduced space
dimension:
\begin{equation}
i \hbar {\partial\over{\partial \tau }}\Psi 
=
\left(-  
{\hbar ^{2}\over{2m}}
\bbox{\nabla }_{\perp}^{2} + U \right)\Psi \ ,
\label{ica}
\end{equation}
This formal analogy allows one to discuss the {\em  stationary}
solutions in terms of the wave moving in the direction $\bbox{v}$, and
call $x$ the current coordinate of the wave. The Feynman path
integral equivalent to Eq.(\ref{ica}) gives an alternative method of
solving the equation.

The probability current $\bbox{J}$ is derived from the standard
expression $\bbox{J}= {\hbar \over{m}} \Im \Psi_{Sch}^{*}\bbox{\nabla
}\Psi_{Sch}$ and the definition Eq.(\ref{mca}). In the main
approximation, the components parallel, $\bbox{J}_{||}$, and
perpendicular, $\bbox{J}_{\perp}$, relative to $\bbox{v}$, are
\begin{equation}
\bbox{J}_{||}= \bbox{v}|\Psi |^{2}\;\; , \;\;  
\bbox{J}_{\perp} = {\hbar \over{m}} \Im
\Psi^{*}\bbox{\nabla}_{\perp}\Psi \ .
\label{jca}
\end{equation}
These expressions are consistent with the current conservation and
Eq.(\ref{hca}) or Eq.(\ref{ica}): Indeed, the continuity equation
which follows from Eq.(\ref{ica}),
\begin{equation}
{\partial |\Psi |^{2}\over{\partial \tau}}+
\text{{\bf div }} \bbox{J}_{\perp}=0
\label{kca}
\end{equation}
 is equivalent to 
$\text{div }\bbox{J}=0$ with $\bbox{J}$ from Eq.(\ref{jca}). The continuity in Eq.(\ref{kca}) means that
the paraxial wave function can be normalized: 
\[
\int
d \bbox{r}_{\perp} |\Psi |^{2} =1
\]
fixing the total flux in the beam  to $\bbox{v}$.

The required solution to the paraxial equation can be chosen by
imposing a proper boundary condition to Eq.(\ref{hca}) (``initial''
condition in the case of Eq.(\ref{ica}))

\paragraph{Paraxial approximation: conditions of applicability} The
paraxial theory is based on the approximation $E(\bbox{p})-
E(\bbox{p}_{0})\approx v\delta p_{||}+ {1\over{2m}}(\delta
\bbox{p}_{\perp})^{2}$, $\delta \bbox{p}\equiv \bbox{p}-\bbox{p}_{0}$
where the term $ {1\over{2m}}(\delta \bbox{p}_{||})^{2}$ is
neglected. This is justifiable if the angle, $\theta \sim {\delta
p_{\perp}\over{p_{0}}}$, between $\bbox{p}$ and $\bbox{p}_{0}$ is
small, $\theta \ll 1$.  If the motion is free, $\delta p_{\perp}\sim
{\hbar \over{w}}$ where $w$ is the width of the beam (defined by the
boundary conditions).  Paraxial approximation is therefore applicable
if the beam is wide in the scale of the wave length $\lambdabar$,
\begin{equation}
w \sim {\delta p_{\perp}\over p_{0}}\gg \lambdabar \ .
\label{pca}
\end{equation}

The scattering by the potential $U$ changes the angle by
$\theta_{\text{scat}}\sim U/E$ . The paraxial approximation requires
the small angle scattering to be dominant, $\theta \sim
\theta_{\text{scat}}\ll 1$, so that the theory is applicable only for
fast particles: $E \gg |U|$.  It is important however, that, as in the
case of the eikonal approximation \cite{LanLif3}, the theory is
applicable beyond the Born approximation: the phase shift $\delta
\varphi \sim {U a\over{\hbar v}}\sim {U\over{E}} {pa\over{\hbar}}$,
$a$ being the thickness of the layer where $U\neq 0$, may be large
even for fast particles.

Unlike the eikonal approximation where only the phase variations are
taken into account, the paraxial theory allows also for the change of
the profile of the beam, {\it i.e.} $|\Psi (\bbox{r})|$ due to
diffraction (or, equivalently, to broadening of the wave packet in the
language of Eq.(\ref{ica})).  If $w$ is a typical size of the
transverse structure defined by either the initial condition or
scattering, the ``diffraction blurring of the image'' happens when the
beam travels the distance $x_{\text{diff}}$,
\begin{equation}
x_{\text{diff}}\sim {w^{2}\over{\lambdabar}} \ .
\label{nca}
\end{equation}
The diffraction length $x_{\text{diff}}$ is the typical distance for
the paraxial approximation while region $x\ll x_{\text{diff}}$ is
described by the eikonal approximation.

Applicability of the approximation at large distances requires further
analysis. The paraxial relation $\delta p_{||}= -
{1\over{2p_{0}}}(\delta \bbox{p}_{\perp})^{2}$ is valid up to a small
correction $(\delta p_{||})_{2}$ due to the neglected quadratic term:
$(\delta p_{||})_{2} \simeq {1\over{2p_{0}}}(\delta p_{||})^{2} $. In
the main approximation, $(\delta p_{||})_{2} \sim
p_{\perp}^{4}/{p_{0}^{3}}$. Although small in comparison with $\delta
p_{||}$, the correction is important at long enough distances; it can
be ignored only if ${(\delta p_{||})_{2} x /{\hbar}} < 1$.  Thus, the
paraxial approximation is reliable if the distance traveled by the
beam is not not too large,
\begin{equation}
x < {w^{4}\over{\lambdabar^{3}}} \sim \left({w\over{\lambdabar }}
\right)^{2} x_{\text{diff}} \ .
\label{oca}
\end{equation}
If the main condition Eq.(\ref{pca}) of applicability of the paraxial
approximation is met, this requirement is distances large compared
with the typical diffraction length $ x_{\text{diff}}$.

\subsection{Paraxial approximation: magnetic scattering}
\label{magn}

In this section the 2D paraxial theory is applied to the case of an
external magnetic field; for simplicity $U=0$, generalization to
$U\neq 0$ is straightforward.

The paraxial wave equation, the gauge covariant form of Eq.(\ref{hca})
with $U=0$, reads
\begin{equation}
\Big(i \hbar v \partial_{x} +{\hbar^{2}\over{2m}}\partial^{2}_{y}   \Big)\psi =0
\label{5a/}
\end{equation}
where $\partial_{x,y}\equiv {\partial\over{\partial x,y}}-i
{e\over{\hbar c}}A_{x,y}$ $\bbox{A}$ being the vector potential; the
$x-$axis is chosen in the direction of the propagation $\bbox{v}$.
The current density is given by Eq.(\ref{jca}) {\it if} modified by
the standard diamagnetic term \cite{LanLif3}.

\begin{figure}[h] 
 \centerline{\epsfig{file=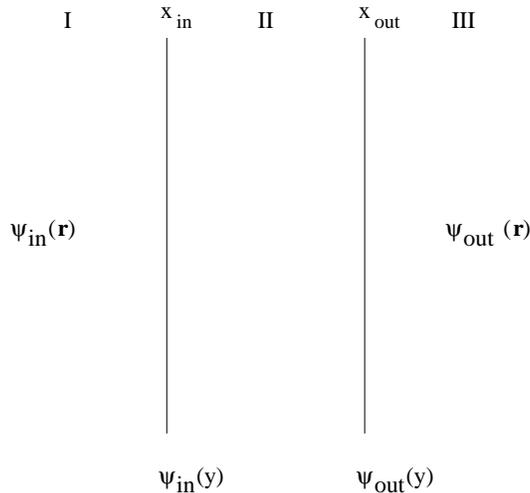,height=0.3\textheight,angle=-90}}
\caption{Magnetic scattering geometry. Initially, the particle moves 
in the field free region $I$; $\psi_{in}(x,y)$ defined by
Eq.(\ref{8d0}) has the meaning of the incoming wave function in the
gauge where the vector potential is zero in $I$.  The particle is
scattered by the magnetic field in the region $II$ and then moves
freely in the region $III$.  In the gauge where the vector potential
is zero in $III$, the outgoing wave is described by
$\psi_{out}(\bbox{r})$.  To solve the scattering problem means to
relate the interface value of the outgoing wave, $\psi_{out}(y)\equiv
\psi_{out}(x_{out},y)$, to that of the incoming one,
$\psi_{in}(y)\equiv \psi_{in}(x_{in},y)$.}
\label{I-II-III}
\end{figure}

It is convenient to consider first the situation when the field is
present only in a finite region.  Divide the space into three regions:
$x<x_{in}$, incoming (I); $x_{in}< x<x_{out}$ scattering (II); and
outgoing region (III), $x> x_{out}$ (see Fig. \ref{I-II-III}).  In
regions I and II a magnetic field is absent.  Present the wave
function in I as
\begin{equation}
\Psi(\bbox{r})=
e^{i{e\over{\hbar}}\int\limits_{\bbox{R_{I}}}^{\bbox{r}} d \bbox{r}
\cdot 
\bbox{A} } \psi_{\text{in}}(\bbox{r}) \;\; , \;\;  x\leq x_{in} \ ,
\label{8d0}
\end{equation} 
and by this define $\psi_{\text{in}}$ .  The integral in
Eq.(\ref{8d0}) does not depend on the path of the integration if the
latter is in the magnetic free region I.  The overall phase of
$\psi_{\text{in}}$ depends on the choice of $\bbox{R}_{I}$ and the
gauge of $\bbox{A}$. It is convenient to put $\bbox{R}_{I}$ on the
I-II interface, $\bbox{R}_{I}= (x=x_{in}, y= y_{*})$, with somehow
chosen $y_{*}$.

The new function $\psi_{\text{in}}$ has the properties of the wave
function in the gauge $\bbox{A}=0$: It obeys the free equation,
\begin{equation}
\Big(i \hbar v {\partial\over{\partial x}} +
{\hbar^{2}\over{2m}}{\partial^{2}\over{\partial y^{2}}}   \Big)\psi_{in}=0
\label{qca}
\end{equation}
and the probability current is given by Eq.(\ref{jca}) ({\em without}
any diamagnetic term).

The wave function of the incoming beam is the input to the scattering
problem: It is assumed that the problem of a free propagation in I is
solved, and the incoming wave at the I-II interface,
$\psi_{\text{in}}(\bbox{r})|_{x=x_{in}}= \psi_{\text{in}}(y)$, is
known.  The normalization condition
\[
\int\limits_{-\infty}^{\infty}dy\,|\psi_{\text{in}}(y)|^{2} =1 \  
\]
makes the total (conserving) flux in the $x$-direction equal to the
velocity $v$.  With the above choice of $\bbox{R}_{I}$, the wave
function Eq.(\ref{8d0}) at the boundary of the region I reads
\begin{equation}
\Psi(x,y)|_{x=x_{in}}= 
 e^{i{e\over{\hbar}}\int\limits_{y_{*}}^{y}
dy_{1}\,A_{y}(x_{in}, y_{1})} 
\psi_{\text{in}}(y)
\ .
\label{4bz}
\end{equation}
where the  integration is performed along a piece of the straight line
$x = x_{in}$.

The wave at $x>x_{in}$ has to be found from the paraxial equation
Eq.(\ref{5a/}), solved with Eq.(\ref{4bz}) as the boundary condition.
The solution in the both, scattering (II) and outgoing (III), regions
may be generally written as,
\begin{equation}
\Psi (x,y) 
=
\int\limits_{-\infty}^{\infty}dy'\ G^{R}(x,y;
x_{in},y')
 e^{i{e\over{\hbar}}\int\limits_{y_{*}}^{y'}
dy_{1}\,A_{y}(x_{in},y_{1})} 
\psi_{\text{in}}(y') \;\; , \;\;  x>x_{in}
\label{7ba}
\end{equation}
where the  Green function, $G^{R}(\bbox{r},\bbox{r}')$ 
solves
\begin{equation}
\Big( i \hbar v \partial_{x} +{\hbar^{2}\over{2m}}\partial^{2}_{y}
\Big)G^{R}(\bbox{r},\bbox{r}')\ = 
i \hbar v \delta (\bbox{r}-\bbox{r}') \;\; , \;\;
G^{R}=0 \text{ for } x< x' \ .
\label{6ba}
\end{equation} 

Similar to the above consideration of I (see Eq.(\ref{8d0})), one
defines in III a new function, $\psi_{\text{out}}$, by
\begin{equation}
\Psi(\bbox{r})=
e^{i{e\over{\hbar}}\int\limits_{\bbox{R_{III}}}^{\bbox{r}} d \bbox{r}
\cdot 
\bbox{A} } \psi_{\text{out}}(\bbox{r}) \;\; , \;\;  x \geq x_{out} \ ;
\label{8ba}
\end{equation}
again, the whole path of integration must be in the field free region
III; choose the initial point of the integration path
$\bbox{R}_{III}=(x_{out}, y_{*})$ (or any other point at the $II-III$
interface).  Analogously to $\psi_{\text{in}}$ in I,
$\psi_{\text{out}}(\bbox{r})$ obeys the free Eq.(\ref{qca}) and is
fully defined in the whole region III by $\psi_{\text{out}}(y)$ that
is the boundary value at the II-III interface,
$\psi_{\text{out}}(y)\equiv \psi (x_{out},y)$. From Eq.(\ref{8ba}) and
(\ref{7ba}), we see that
\begin{equation}
\psi_{\text{out}}(y) =
\int\limits_{-\infty}^{\infty}dy'\ 
{\cal G}(x_{out},y; x_{in},y')
\psi_{\text{in}}(y')
\label{9ba}
\end{equation}
where the Green function ${\cal G}$ is
\begin{equation}
{\cal G}^{R}(x,y;x',y')\equiv 
e^{-i{e\over{\hbar}}\int\limits_{y_{*}}^{y}
dy_{2}\,A_{y}(x,y_{2})}
G^{R}(x,y; x',y')\ 
 e^{i{e\over{\hbar}}\int\limits_{y_{*}}^{y'}
dy_{1}\,A_{y}(x',y_{1})}  \; .
\label{0ba}
\end{equation}
Eq.(\ref{9ba}) combined with Eq.(\ref{0ba}) relates the outgoing wave
amplitude $\psi_{\text{out}}$ to the incoming wave $\psi_{\text{in}}$,
and thus gives a general solution to the magnetic scattering problem
in the paraxial approximation.

Known  $\psi_{\text{out}}$, one finds $\psi_{\text{out}}(x,y)$ in the
outgoing region $III$ as 
\begin{equation}
\psi_{\text{out}}(x,y)= \int\limits_{- \infty}^{\infty}dy' 
G^{R}_{0}(x-x_{out},y-y')\psi_{\text{out}}(y')\;\; , \;\;  x \geq x_{out}
\label{rca}
\end{equation}
with
\begin{equation}
G^{R}_{0}(x,y)=  \theta (x)
{1\over{\sqrt{2 \pi i \lambdabar x}}}\,
 e^{{i\over{2 \lambdabar x}} y^{2}} \ ,
\label{sca}
\end{equation}
being the free propagator \cite{footnote99}.

\subsection{The eikonal  approximation} 
\label{eikonal}

If the scattering region II is narrow enough compared with the
diffraction length Eq.(\ref{nca}), one can neglect the transverse
derivatives in Eq.(\ref{5a/}). This limit corresponds to the
well-known eikonal approximation \cite{LanLif3}. The wave function
obeys the eikonal equation ($U=0$)
\begin{equation}
i \hbar \partial_{x} \psi =0 \ .
\label{uca}
\end{equation}

Solution to Eq.(\ref{6ba}) reads in the eikonal limit
\begin{equation}
G^{R}(x,y; x',y') = 
e^{i{e\over{\hbar}}\int\limits_{x'}^{x}dx''\,A_{x}(x'',y)}
\delta (y-y')\; \theta (x-x') \ .
\label{dca}
\end{equation}

\begin{figure}[h] 
\centerline{\epsfig{file=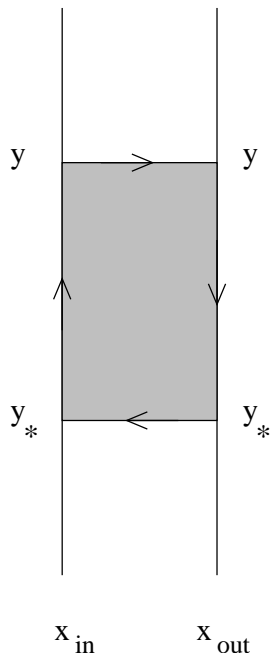,height=0.15\textheight,angle=-90}}
\caption{The eikonal solution to the magnetic scattering problem. 
The scattering amplitude (see Eq.(\ref{eca})), is controlled by the
flux encircled by the contour shown in the picture.
}
\label{eiko}
\end{figure}

Substituting Eq.(\ref{dca}) into Eqs.(\ref{0ba}) and (\ref{9ba}), one
gets the eikonal solution to the magnetic scattering problem,
\begin{equation}
\psi_{\text{out}}(y)= 
e^{i {e\over{\hbar c}}\Phi (y)} \psi_{\text{in}}(y)
\label{eca}
\end{equation} 
where $\Phi (y)$ is the total magnetic flux through the directed area
(see Fig. \ref{eiko}) restricted by the path $(x_{in},
y_{*})\rightarrow (x_{in}, y)\rightarrow (x_{out}, y))\rightarrow
(x_{out}, y_{*})$; choice of $y_{*}$ is arbitrary affecting only the
overall phase of $\psi_{\text{out}}$ \cite{footnote2}.

\section{Illustrations: Aharonov-Bohm scattering}
\label{AB}

To illustrate the usage of the paraxial theory, the scattering on the
Aharonov-Bohm magnetic line is considered in this section as an
example. Some results presented in this section has been published in
a short communication Ref.\cite{She98}.

The scattering problem is formulated as follows.  The incident wave
$\psi_{in}= \psi_{\text{in}}(y)$ comes from the negative side of the
$x$-axis and the charge sees a magnetic line (extending in the
$z$-direction), which creates the magnetic field $b(x,y)= \Phi_{0}
\delta (x) \delta (y)$.  The final goal is to characterize the
outgoing wave at $x>0$.

As the magnetic field is finite only at the origin, the scattering
region II in the terminology of Section \ref{magn} can be shrunk to
just the line $y=0$, so that $x_{in}= - 0$ and $x_{out}= + 0$.  Seeing
that scattering region is narrow, the eikonal approximation discussed
in Section \ref{eikonal} is applicable, except, perhaps, in the
immediate vicinity of the singular magnetic line point $y=0$.  The
flux function, $\Phi(y)$, in the eikonal expression Eq.(\ref{eca}) is
easily found to be $\Phi(y)= \Phi \theta (y)$ (if $y_{*}$ is chosen at
$-\infty $) .  
The
outcoming wave  reads
\begin{equation}
\psi_{\text{out}}(y) = \exp \left( 
-i \pi 
\tilde{\Phi}
sign (y)
\right)  \psi_{\text{in}}(y) \; . 
\label{vca}
\end{equation}
where $\tilde{\Phi}\equiv {\Phi \over{\Phi_{0}}}$ \cite{ffoot}.  Same
result can readily obtained \cite{She98} by solving the first order
differential equation Eq.(\ref{uca}) with the vector potential in the
gauge
\[
A_{x}(x,y)= -{1\over 2}  \Phi \, sign(y)\, \delta(x)  
\;\; , \;\;  A_{y}(x,y) =0 \; .
\]

The freely propagating outgoing wave is found from Eq.(\ref{rca}) with
$x_{out} =0$ and $\psi_{\text{out}}$ from Eq.(\ref{vca})
Ref.\cite{She98},
\begin{equation}
\psi_{\text{out}}(x,y) = \psi_{\text{in}}(x,y)
\cos{\pi\tilde{\Phi}}+ i V(x,y)
\sin{\pi\tilde{\Phi}}
\;\; , \;\;  x> 0 \ ,
\label{2ca}
\end{equation}
where 
$ \psi_{\text{in}}(x,y)$ 
is the incoming wave continued to the region $x>0$,
\begin{equation}
 \psi_{\text{in}}(x,y) =  
\int\limits_{- \infty}^{\infty}dy' 
G^{R}_{0}(x,y-y')\psi_{\text{in}}(y') \ ,
\label{3ca}
\end{equation}
{\it i.e.}  the wave in the absence of the Aharonov-Bohm line,
  and  
\begin{equation}
 V(x,y)= 
\int\limits_{- \infty}^{\infty}dy' 
G^{R}_{0}(x,y-y')\, \hat{y'}\,\psi_{\text{in}}(y') \ .
\label{4ca}
\end{equation}
Eqs.(\ref{2ca} \ref{4ca}) give the solution of the Aharonov-Bohm
problem in the paraxial approximation for an arbitrary incoming wave
$\psi_{in}(y)$.

Using Eq.(\ref{sca}), one can check that
\begin{eqnarray}
\int\limits_{- \infty }^{\infty } dy\, |\psi_{\text{in}}(x,y)|^{2}
 =   1 
\;\; &,& \;\;  
\int\limits_{- \infty }^{\infty } dy\, |V(x,y)|^{2}
=  1 \ ,
\nonumber 
\\   
\Im \int\limits_{- \infty }^{\infty } dy \;
\psi^{*}_{\text{in}}(x,y&)&V(x,y)     =   0  ;
\nonumber 
\end{eqnarray}
for any $x>0$ and arbitrary (normalized) $\psi_{\text{in}}$; these
relations guarantee the current conservation \cite{footnote11}.

The solution in Eq.(\ref{2ca}) gives a convenient tool for studying
interaction of beams (waves packets) with the Aharonov-Bohm flux line.
Some examples are considered below.

\subsection{Plane incident wave}

\begin{figure}[b]
\centerline{\epsfig{file=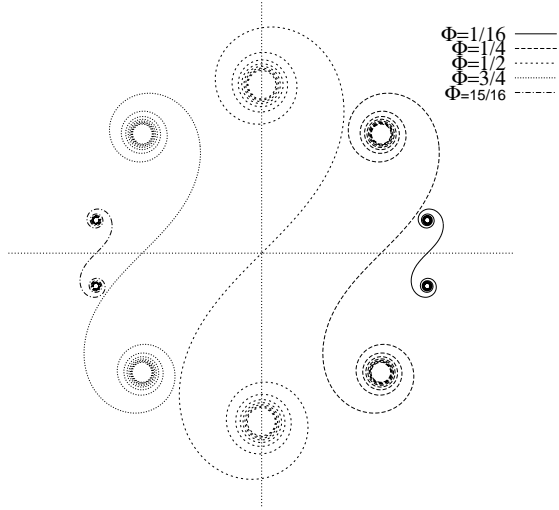,height=200pt,angle=0}}
\caption{
The paraxial
wave-function Eq.(\ref{dda}) on the $\psi $ -complex plane for the
fluxes 
${1\over 16} \Phi_{0}$, 
${1\over 4} \Phi_{0}$, 
${1\over 2} \Phi_{0}$, 
${3\over 4} \Phi_{0}$, 
${15\over 16} \Phi_{0}$.
The points on the spirals are uniquely characterized by the spiral
arc-length counted from the point where $Im \, \psi =0$ and measured
in units proportional to $\sin \tilde{\Phi }$.  For the point with the
coordinates $x,y$, the wave function $\psi (x,y)$ corresponds to the
point on the Cornu spiral with the arc-length equal to $\tilde{y}=
y/\protect\sqrt{2 \lambdabar x}$ . In accordance with
Ref.\protect\cite{BerShe99}, the exact Aharonov-Bohm solution can be
also mapped to the Cornu spiral; for a general scattering angle the
mapping condition is that arc-length $ \protect\approx\protect
\sqrt{(r-x)/\lambdabar }$ in proper units.  }  \label{cornu}
\end{figure}

Plane wave corresponds to $\psi_{\text{in}}(y)= 1$.  From
Eqs.(\ref{3ca}) and (\ref{sca}), one gets $\psi_{\text{in}}(x,y)= 1$,
and from Eq.(\ref{4ca}) Ref.\cite{She98},
\begin{equation}
V(x,y)= K (\tilde{y}) \;\; , \;\;  \tilde{y}= {y\over{\sqrt{2 \lambdabar x}}}
\label{ada}
\end{equation}
where $K(t)$ is a familiar function,
\begin{equation}
K(t) =
{2\over{\sqrt{i \pi  }}}\,
\int\limits_{0}^{t}\, dt'\,
 e^{i t'^{2}} \ ,
\label{cda}
\end{equation}
related to the Fresnel integrals.
The outgoing wave Eq.(\ref{2ca}) reads
\begin{equation}
\psi_{\text{out}}(x,y) = 
\cos{\pi\tilde{\Phi}}+ i K(\tilde{y})
\sin{\pi\tilde{\Phi}} \ .
\label{dda}
\end{equation}
The complex wave function $\psi_{\text{out}}(x,y)$ is a function of
the real parameter $\tilde{y}$ only. This means, that
$\psi_{out}(x,y)$ for any point of the half-plane $x>0$ spans a {\em
line} on the complex plane $\Re \psi - \Im \psi$. This line
(Fig. \ref{cornu}) is the Cornu spiral well-known in the diffraction
theory \cite{BorWol59}.  The property of the Cornu spiral is that
$(dt)^{2}$ is proportional to $(d \ell)^{2}$ where $d \ell $ is the
distance between the points corresponding to $t$ and $t+dt$. From
Eq.(\ref{ada}) and Eq.(\ref{cda}), the element $d \ell$ of the length
along the spiral (the arc-length) and $d \tilde{y}$ are related as $(d
\ell)^{2}= {4\over \pi } \sin^{2}(\pi \tilde{\Phi})(d \tilde{y})^{2}$.
Since the wave function Eq.(\ref{dda}) is real for $\tilde{y}=0$, the
arc-length counted from the spiral point with $\Im \psi =0$ is
proportional to $\tilde{y}$

The singular Aharonov-Bohm line problem gives a rather tough test for
the validity of the paraxial approximation.  The evaluation of its
accuracy has been made simple by the recent observation
Ref.\cite{BerShe99} that the exact solution can be presented in the
form Eq.(\ref{dda}) with $\tilde{y} \rightarrow \tilde{Y}$ with
$\tilde{Y}$ found from
\begin{equation}
 \tilde{Y}^{2} =  {(r-x)\over \lambdabar }  + 
\varphi \left({1\over 2}- \tilde{\Phi} \right) + \ldots
\; ,
\label{ejb}
\end{equation}
 $r$ and $\varphi $ being the cylindrical coordinates; the terms not
shown in Eq.(\ref{ejb}) are of order ${\cal O}\left({\lambdabar \over
r} \right)$ \cite{BerShe99}.  (Unexpectedly, the Aharonov-Bohm wave
function on the $x-y$ {\em plane} can be mapped on the Cornu spiral
not only in the forward direction but for any scattering angle (and $r
\gg \lambdabar $), see Ref. \cite{BerShe99} for details.)  The
paraxial approximation gives correctly the leading term
$\sqrt{{r-x\over \lambdabar }} \approx\tilde{y}$ in the vicinity of
the forward direction and short wave length $\lambdabar \ll r$.

\subsection{Finite size beam}

The scattering of a plane wave by the Aharonov-Bohm line is highly
singular: Eq.(\ref{dda}) shows that in the forward direction, $\varphi
\ll \sqrt{{\lambdabar \over x}}$ {\it i.e.}  $\tilde{y}\ll 1$, the
wave function equals to $\cos \pi \tilde{\Phi}$ and does not converge
at large distances to the plane incident wave as assumed in the
standard scattering theory.  Known from the exact solution
\cite{Ber80}, this anomaly is the reason why the text-book scattering
theory fails: The incoming plane wave and scattered wave cannot be
separated \cite{Ber80} and the scattering amplitude cannot be
introduced as the object carrying the complete information about
scattering \cite{She98}.

The singular behaviour is due to the combination of two factors: (i)
the infinitely long range of the interaction with the line; (ii) the
infinite extension of the plane wave.  Since any physical state has a
finite transverse extension $W$, the behaviour of the potential beyond
the width $W$ is irrelevant, and the singularities are expected to be
regularized.

\begin{figure}[h] 
 \centerline{\epsfig{file=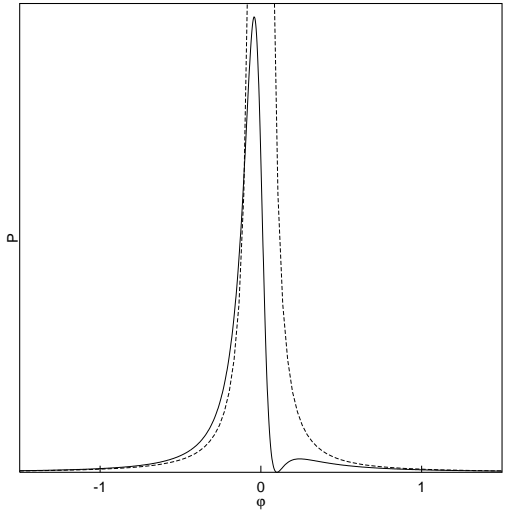,height=0.3\textheight,angle=-90}}
\caption{
Scattering of a beam by the Aharonov-Bohm line.  The solid line shows
the angular distribution of the intensity $P(\varphi )$ in
Eq.(\ref{6hb}) for the flux $\tilde{\Phi }= +1/4$ and the beam angular
width $\varphi_{0}= 0.1$. The dashed, line which is the Aharonov-Bohm
cross-section Eq.(\ref{8hb}) fits well the distribution at large
enough angles $\gg \varphi_{0}$ where the intensity can be attributed
to scattering.  The central, $|\varphi|\protect\lesssim \varphi_{0}$,
peak has a nontrivial structure, asymmetric relative to $\varphi
\rightarrow - \varphi $.  The small angle left-right asymmetry, which
is absent in the scattering cross-section, is responsible for the
effective Lorentz force exerted by the Aharonov-Bohm line.}
\label{PvsFi} \end{figure}

To show this, suppose that the incoming wave is beam-like with the
profile
\[
\psi_{in}(y) =  e^{- {y\over |W|}} \; ,
\]
where $W $ is the beam width; it is assumed that $W \gg \lambdabar $
so that the paraxial approximation is applicable.  The beam has a
small but finite angular width $\varphi_{0}= \lambdabar/W$.  One can
easily see from Eq.(\ref{2ca}) that at large distances $x\gg W^{2}/
\lambdabar$, the outgoing wave behaves like a spherical wave {\it
i.e.}  $|\psi_{out}(x,y)|^{2} \approx P(\varphi )/ x$ where $\varphi =
y/x$ is the scattering angle and $P(\varphi )$,
\begin{equation}
P(\varphi )= 
{2 \lambdabar  \over \pi   } 
{(\varphi \,\sin \pi \tilde{\Phi} -
\varphi_{0}\cos \pi \tilde{\Phi})^{2}\over 
(\varphi^{2}+ \varphi_{0}^{2})^{2}} \; ,
\label{6hb}
\end{equation}
has the meaning of the angular distribution of the intensity in the
outgoing wave.  As expected, the distribution shown in
Fig. \ref{PvsFi} is perfectly smooth.

For the angles larger than the beam angular width, {\it i.e.} 
$|\varphi | \gg \varphi_{0}$, one gets from Eq.(\ref{6hb}) that
\begin{equation}
P(\varphi )\approx
{2 \lambdabar  \over \pi   }\; 
{\sin^{2}\pi \tilde{\Phi}\over \varphi^{2}} \; ,
\label{8hb}
\end{equation}
recovering the small angle asymptotics of the Aharonov-Bohm
scattering cross-section \cite{AhaBoh59}
\begin{equation}
\left({d \sigma\over d \varphi }\right)_{\!\text{AB}}= {\lambdabar \over 2 \pi}
\,
{\sin^{2}\pi{\tilde{\Phi}}\over \sin^{2}{\phi \over 2 } }
\; .
\label{7hb}
\end{equation}

It would be rather trivial if the angular broadening of the incident
wave led just to a regularization of the forward scattering
singularity. Most important is that a qualitatively new feature
becomes seen: Unlike the Aharonov-Bohm cross-section Eq.(\ref{7hb})
the angular distribution in Eq.(\ref{6hb}) is left-right
asymmetric. The antisymmetric part is concentrated in the forward
direction $|\varphi | \lesssim \varphi_{0}$, {\it i.e.} within the
angular width of the incoming wave {\it i.e.}  in the region where the
scattered and incident waves cannot be separated
\cite{BerChaLar80}. The asymmetry means that the beam is deflected by
the Aharonov-Bohm line as a whole \cite{She98,Ber99}.  The order of
magnitude of the deflection is the initial angular width $\varphi_{0}
\sim \lambdabar /W$.  One may say that the Aharonov-Bohm line not only
scatters the incident wave but also modifies the unscattered part of
the wave. Later, it will be shown that in a system of many
Aharonov-Bohm lines, the deflections by individual lines coherently
add together, and the lines act as an effective magnetic field

The deflection of the beam $\Delta \varphi $ can be presented via the
momentum $\Delta p_{\perp} = {\hbar \over \lambdabar } \Delta \varphi
$ transfered to the charge in the direction perpendicular to its
initial velocity. A calculation details of which are collected in
Appendix \ref{trans} gives the following result Ref.\cite{She98}:
\begin{equation}
\Delta p_{\perp} = \hbar  |\psi_{\text{in}}(0)|^{2} \sin 2\pi{\Phi\over{\Phi_{0}}}
\label{0hb}
\end{equation}
here the momentum transfer is expressed via the value of the {\it
normalized} incoming wave at the position of the line.  By comparison
with the exact theory, the validity of this paraxial result has been
confirmed by Berry \cite{Ber99}.

\section{Density matrix}
\label{dens}

The theory of magnetic scattering presented in Section \ref{magn} is
gauge invariant only in a limited sense. Although Eq.(\ref{9ba}) holds
in arbitrary gauge, each of the objects there, $\psi_{\text{in}}$,
$\psi_{\text{out}}$, and ${\cal G}$ is gauge dependent, although only
through the overall phase.  For example, under $\bbox{A}\rightarrow
\bbox{A} + \bbox{\nabla }\chi$ simultaneously with $\Psi \rightarrow
e^{i {e\over{\hbar c}} \chi}$, the incoming wave $\psi_{\text{in}}$
Eq.(\ref{8d0}) is modified as $\psi_{\text{in}}\rightarrow
e^{i{e\over{\hbar}}\chi(\bbox{R}_{I})}\psi_{\text{in}}$; we see also
that the overall phase depends on the arbitrarily chosen
$\bbox{R}_{I}$.  Of course, the observables are independent from the
global phase and the above gauge dependence does not create any
problem.

A truly gauge invariant theory can be formulated 
in terms of the  
 by-linear in $\Psi $ and $\Psi^{*}$ ``density matrix'' $\rho (y_{1},y_{2}; x)$ 
 defined as
\begin{equation}
\rho (y_{1},y_{2}; x)\equiv  
e^{-i{e\over{\hbar}}\int\limits_{y_{2}}^{y_{1}}dy'\,A_{y}(x,y')}
\Psi (x, y_{1}) \Psi^{*}(x, y_{2}) \ .
\label{aca}
\end{equation}
The  ``density matrix'' $\rho$ 
carries  the full  quantum 
information needed to find observables and is gauge invariant. 
In the field free regions, the density matrix is built from
$\psi_{\text{in}}$ and $\psi_{\text{in}}^{*}$ or
$\psi_{\text{out}}$ and $\psi_{\text{out}}^{*}$; these combinations 
depend on neither  gauge nor $\bbox{R}_{I,II}$.

The current density the $x$- and $y$- directions are (cf. Eq.(\ref{jca}))
\[
J_{x}(x,y)= v \rho (y,y;x) \;\; , \;\;  
J_{y}(x,y)= {\hbar \over{2im}} \left({\partial\over{\partial y_{1}}}
- {\partial\over{\partial y_{2}}}  
\right)
\rho(y_{1}, y_{2}; x) \rule[-1.5ex]{.1ex}{3ex}_{y_{1}= y_{2}=y}
\]

Seeing that the evolution of the wave function $\Psi $ in
Eq.(\ref{aca}) is given by the propagator $G^{R}$ Eq.(\ref{6ba}),
the density matrix evolves 
from $x'$ to $x$ ($x>x'$) 
as
\begin{equation}
\rho (y_{1},y_{2};x)= 
\int\limits_{-\infty}^{\infty} 
dy_{1}' \, dy_{2}'\, 
{\cal G}^{R}(y_{1},x; y_{1}',x' )\,
\rho (y_{1}',y_{2}';x')\,
{\cal G}^{A}(y_{2}',x'; y_{2},x )
\label{bca}
\end{equation}
where
${\cal G}^{R}$ is defined by  (\ref{0ba}) and the advanced
Green function ${\cal G}^{A}$
\[
{\cal G}^{A}(\bbox{r}_{1},\bbox{r}_{2})= \left({\cal
G}^{R}(\bbox{r}_{2},\bbox{r}_{1})\right)^{*} \ .
\]

Introducing the two-particle Green function
\begin{equation}
{\cal K}(y_{1},y_{2};x| y_{1}',y_{2};,x')=
{\cal G}^{R}(y_{1}, x| y_{1}', x'){\cal G}^{A}(y_{2}', x'| y_{2}, x) \ ,
\label{qda}
\end{equation}
Eq.(\ref{bca}) can be written as 
\begin{equation}
\rho (y_{1},y_{2};x)= 
\int\limits_{-\infty}^{\infty} 
dy_{1}' \, dy_{2}'\, 
{\cal K}(y_{1},y_{2};x| y_{1}',y_{2};,x')
\rho (y_{1}',y_{2}';x')\,
\label{6da}
\end{equation}

As before, the incoming wave enters the  scattering problem as the boundary
condition at $x= x_{in}$:  $\rho(y_{1}, y_{2}; x_{in})=
\rho_{in}(y_{1},y_{2})$. 
Further propagation of the incoming beam is given by
Eq.(\ref{6da})
with $x'= x_{in}$.

For future references, we note the following property of the Green function: 
\begin{eqnarray}
\int\limits_{- \infty }^{\infty }d y
\, G^{R}(x,y; x', y_{1}) 
\,G^{A}( x', y_{2}; x,y)
     & =  & \theta (x-x')\, \delta (y_{1}-y_{2}) \ ,  \label{5ca1}           \\
\int\limits_{- \infty }^{\infty }d y
\, G^{R}(x,y_{1}; x', y) 
\,G^{A}( x', y ; x,y_{2})
     & =  & \theta (x-x') \,\delta (y_{1}-y_{2}) \ ,                
\label{5ca}
\end{eqnarray}
this relations expresses the current conservation \cite{footnote6ca}.

In particular, from Eq.(\ref{5ca1}), one gets the conservation of the
total current in the beam:  
\[
\int dy\  \rho (y, y; x)=
\int dy' \rho (y', y'; x=0)
\]

\subsection{Path integral representation}
\label{path}

Similar to Eq.(\ref{ica}), the stationary paraxial equation in 2D may
be mapped to a time dependent 1D problem:
\begin{equation}
i \hbar {\partial\over{\partial \tau }}\psi = \left(
-{\hbar^{2}\over{2m}}\left({\partial\over{\partial y}}-i
{e\over{\hbar c}}a \right)^{2} + e \varphi   \right)\psi =0
\label{4ba}
\end{equation}
where ``time'' $\tau = {x\over{v}}$, $a= A_{y}$, and $\varphi = - {v\over{c}}A_{x}$.  
In the effective 1D problem, the particle  moves 
in the  ``electric field'', 
\begin{equation}
F= - {1\over{c}} \dot{a} - \nabla \varphi \ , 
\label{tca}
\end{equation}
defined by the ``vector potential'', $a$, and the ``scalar
potential'', $\varphi$.  The gauge transformation,
$\bbox{A}\rightarrow \bbox{A} + \bbox{\nabla }\chi$, translates to $a
\rightarrow a + \nabla \chi$, $\varphi \rightarrow \varphi -
{1\over{c}} \dot{\chi}$, so that the effective electric field
Eq.(\ref{tca}) is indeed gauge invariant.  Locally, the electric field
is related to the magnetic field of the original problem $ b(x, y)$ as
$F= -v b$.

The mapping to the effective 1D problem allows one to use a convenient
 path integral representation for the paraxial
propagators. Obviously, $G^{R}$ is just the retarded Green function
for nonstationary equation Eq.(\ref{4ba}) and in the Feynman path
integral representation
\[
G^{R}(x,y;x',y')= 
\int\limits_{y'}^{y} {\cal D}[y(x )]
e^{iS[y(x )]}
\]
where the action of the effective 1D problem
$S= {1\over{\hbar}}\int d \tau  (m \dot{y}^{2}/2+ e a\dot{y}/c
- e \varphi )$ 
translates as
\begin{equation}
S[y(x )]= {1\over{2\lambdabar }}\int\limits_{x'}^{x}dx\, 
 y^{ 2}_{x}+ {e\over{\hbar c}} \int\limits_{y=y(x)}\,d\bbox{r\cdot A}
 \ ,
\label{lda} 
\end{equation}
where $y_{x}= {dy\over{dx}}$. The action corresponding to ${\cal
G}^{R}$ differs from Eq.(\ref{lda}) in the path of integration which
should be extended in an obvious way to include the additional
exponential factors in the definition of ${\cal G}^{R}$
Eq.(\ref{0ba}).

\begin{figure}[t]
\centerline{ 
\epsfig{file=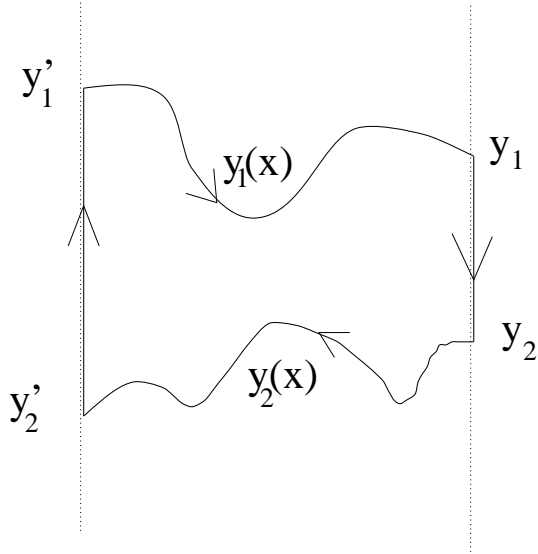,height=200pt,angle=-90}
}
\caption{The integration in Eq.(\ref{jda}) is performed with respect
to the paths $y_{1}(x)$ and $y_{2}(x)$. Due to the phase factors
introduced in Eqs.(\ref{8d0}), and (\ref{8ba}),
the action contains the vector potential integrated along the closed
loop shown here.} 
\label{curly-path}
\end{figure}

With the help of Eq.(\ref{lda}), the two particle Green function
Eq.(\ref{qda}) can be presented as
\begin{equation}
{\cal K}(y_{1},y_{2};x| y_{1}',y_{2};,x')= 
\int 
{\cal D}[y_{1}(x)]
{\cal D}[y_{2}(x)]
e^{i
{\cal S}[y_{1}(x),y_{2}(x)]
} \ ,
\label{jda}
\label{nda}
\end{equation}
where 
\[
{\cal S}= {\cal S}_{0}+ { 2\pi\over{\Phi_{0}}}
\Phi ([y_{1}],[y_{2}]) 
\ ,
\]
${\cal S}_{0}$ being
the free motion contribution,
\begin{equation}
{\cal S}_{0}[y_{1}(x),y_{2}(x)] = 
{1\over{2 \lambdabar }}
\int\limits_{x'}^{x}dx\, 
\left(
 y_{1x}^{  2}-
 y_{2x}^{2}
\right) \ ,
\label{oda}
\end{equation}
and $\Phi $ is the flux
\begin{equation}
\Phi ([y_{1}],[y_{2}]) 
=
\oint\limits_{C([y_{1}],[y_{2}])} d \bbox{r\cdot A}
\label{pda}
\end{equation}
threading the (oriented) area bounded by the paths $y_{1}(x)$ and
$y_{2}(x)$ and the vertical lines at $x$ and $x'$ (see
Fig.\ref{curly-path}).

The path integral representation is used below to perform averaging
with respect to the gauge field.

\section{Gaussian Random magnetic field}
\label{random}

This Section concerns the averaging the two-particle Green's function
Eq.(\ref{qda}) with respect to the Gaussian random magnetic field
Eq.(\ref{vhb}).  In the path integral representation Eq.(\ref{jda}),
the random field enters via the flux $\Phi([y_{1}], [y_{2}])$
Eq.(\ref{pda}).  For the Gaussian field Eq.(\ref{vhb}), the averaged
value
\[ 
\left< 
\exp \left( 2\pi i{\Phi([y_{1}], [y_{2}])\over{\Phi_{0}}}
\right)\right> 
= \exp \left(  - {\cal A}_{\text{no}}([y_{1}-y_{2}])\over{2 {\cal L}^{2}}\right)
\]
where ${\cal A}_{\text{no}}$ 
\[
{\cal A}_{\text{no}}([y_{1}-y_{2}])=
\int\limits_{x'}^{x}dx\,|y_{1}-y_{2}| 
\]
is the non-oriented area bounded by the paths $y_{1}(x)$ and
$y_{2}(x)$ and lines $x=x'$ and $x=x$ (see Fig.\ref{curly-path}).
Further calculations are rather simple thanks to the fact that $\left<
e^{i 2\pi{\Phi\over{\Phi_{0}}}}\right>$ is a functional only of
$y_{1}-y_{2}$. This important simplification is a property of the
models with a $\delta $-correlated magnetic field.

In the variables
\[
y= y_{1}-y_{2} \;\; , \;\;  Y = {1\over{2}}\left(y_{1}+y_{2} \right) 
\ ,
\]
the kinetic energy contribution
${\cal S}_{0}$ Eq.(\ref{oda}) reads  after integration by parts, 
\[
{\cal S}_{0}[y_{1}(x),y_{2}(x)] = 
{1\over{2 \lambdabar }} y_{x} Y |_{x'}^{x}-
{1\over{2 \lambdabar }}
\int\limits_{x'}^{x}dx\, 
y_{xx}Y \ .
\]
Since $Y(x)$ enters only the ${\cal S}_{0}$, the  integration
$e^{i{\cal S}}$ with respect to $Y$ gives $\delta
(y_{xx})$. This means, that the integration with respect to $y(x)$
is limited to the path with 
$y_{xx}\equiv {d^{2}y\over{dx^{2}}}=0$ that is the straight line
connecting the initial and final points. After this, the integral is easily
calculated.

\begin{figure}[h]
\centerline{ 
\centerline{\epsfig{file=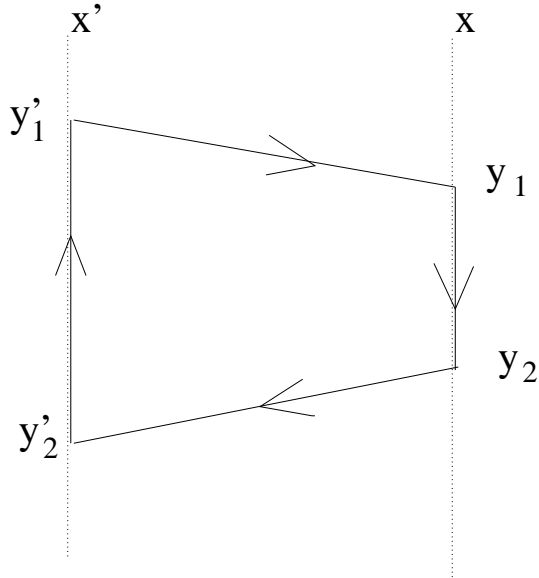,height=200pt,angle=-90}}
}
\vspace{0.5cm}
\caption{The straight line trajectories contributing to the Feynman
integral.  ${\cal A}_{\text{no}}$ in Eq.(\ref{vda}) is the geometrical
area inside the closed loop.  If $(y_{1}-y_{2})(y_{1}'-y_{2}')<0$, the
trajectories cross each other, and the area is built of two
triangles.}
\label{straight}
\end{figure}

Finally, ${\cal K}_{\text{av}}$ that is 
the paraxial two-particle Green function Eq.(\ref{qda}) averaged with
respect to the fluctuation magnetic field, reads
\begin{equation}
{\cal K}_{\text{av}}(y_{1},y_{2};x| y_{1}',y_{2};,x')=
{\cal K}_{\text{0}}(y_{1},y_{2};x| y_{1}',y_{2};,x')\exp \left(- {{\cal
A}_{\text{no}}\over{{2\cal L}^{2}}}\right) \ ,
\label{vda}
\end{equation}
here ${\cal K}_{0}$ is the free two-particle propagator, and ${\cal A}$
is the 
(non-oriented)
area
formed by the straight line trajectories Fig.\ref{straight}, 

\begin{eqnarray}
 {\cal A}    & =  & {1\over 2 } (x-x') \times
\left\{
\begin{array}{lcr}
 |y_{1}+ y_{1}' -
y_{2}- y_{2}'|  \; \;&, &  \; \;  \text{if }
(y_{1}-y_{2})(y_{1}'-y_{2}')>0 \rule[-3ex]{0ex}{0ex} \nonumber\\ 
{(y_{1}-y_{2})^{2}+(y_{1}'-y_{2}')^{2}\over{\mid y_{1}-y_{2} \mid +
\mid y_{1}-y_{2} \mid}}
 &,& \text{if } (y_{1}-y_{2})(y_{1}'-y_{2}')<0
\end{array}
\right.
\label{eea}
\end{eqnarray}

In the variables $y,Y$ and $(x-x')\rightarrow x$, an expanded version
of Eq.(\ref{vda}) reads
\begin{eqnarray}
 K_{\text{av}}(y, Y; y', Y'; x )      &= &
{1\over{2 \pi x \lambdabar}}
\exp \Big[{i\over{x \lambdabar}}(Y-Y') (y-y')
-{1\over{8}} {x\over{{\cal L}^{2}}} \Big(\mid y \mid + \mid y'
\mid 
+ {(y+y')^{2}\over{\mid y \mid}+\mid y' \mid}\Big)  \Big]
\label{ad9}
\end{eqnarray} 

The two-particle Green function in Eq.(\ref{ad9}) allows one to find
{\em averaged} over the random field ``evolution'' of the density
matrix and thus describes correlations in the gauge invariant
observables like the density or the current. For instance, it
describes transmission through a slab with a random magnetic field.
In what follows, Eq.(\ref{ad9}) is applied to some simple cases: (i)
focusing of a coherent wave; (ii) the scattering of the partially
coherent spatially uniform incoming wave.
  
\subsection{Coherent propagation: Focusing}\label{focus}

Let the incident wave $\psi_{in}(y)$  be 
 a converging 
Gaussian beam {\it i.e.} 
\begin{equation}
\psi_{in}(y) = {1\over{(\pi )^{1/4} \sqrt{w}}}
e^{-{y^{2}\over{2}}({1\over{w^{2}}}+ {i\over{\lambdabar f}})}
\; ,
\label{id}
\end{equation} 
where $f$ is the distance to the focal point and $w \ll f$ is the
width of the beam. It is well-known that the ``spherical'' wave like
that in Eq.(\ref{id}) will converge at the focal point $x=f$ producing
a diffraction limited spot with the waist $\sim \lambdabar /\theta $
where $\theta = w/f \ll 1$ is the angular size of the beam as seen
from the focal point.

The distribution of the averaged intensity for the beam propagating in
a random magnetic field can be found with the help of Eq.(\ref{ad9}).
Taking for simplicity only the points on the beam axis $y=0$, the
intensity $ I(x)\equiv \rho (0,0; x)$, reads
\[
I(x)= {1\over{\sqrt{\pi }w}}{1\over{\lambda x}}
\int\limits_{-\infty}^{\infty} \; dY \; dy
e^{i {1\over{ \lambdabar}}\big({1\over{x}}- {1\over{f}}
\big)yY - {Y^{2}\over{w^{2}}}
- {y^{2}\over{4w^{2}}}
- {x\over{4 {\cal L}^{2}}}\mid y \mid } \; .
\]
At the focal point $x= f$, 
\[
I(f)= {1\over{\lambda f}}
\int\limits_{-\infty}^{\infty} \; dy
e^{
- {y^{2}\over{4w^{2}}}
- {f\over{4 {\cal L}^{2}}}\mid y \mid  }\; ,
\]
In the limiting cases, 
\begin{equation}
I(f)=
\left\{
\begin{array}{lcl}
 {2 \sqrt{\pi }\over{\lambda f}}\times w  &\; \; \; \;  , \; \; &
f w \ll {\cal L}^{2}     \\
{8\over{\lambda f}}\times {{\cal L}^{2}\over{f}}
&\; \; \; \; , \; \;    &f w \gg {\cal L}^{2}
\end{array} 
\right. \; .
\label{wr}
\end{equation}
For the conditions when the magnetic field is not important, the upper
line gives the usual diffraction limited value of the intensity; the
width of the spot at the focal plane (line) is of order of $\sim 1/
I(f) \sim \lambdabar / \theta $, where $\theta = w/ f $.  The larger
the aperture $w$, the larger the intensity in the focus and the
smaller the size of the spot.  However, in the presence of the random
magnetic field, the intensity saturates when the aperture $w \sim
{\cal L}^{2}/ f$ and $\theta \sim \left({\cal L}/f \right)^{2}$.

This behaviour is very different from that when the scattering is due
to a random scalar potential with a short correlation length. In this
case, the relevant parameter characterizing the disorder is the focal
length $f$ in units of the mean free path $l$ rather then the size of
the aperture: On the background created by incoherent scattering, one
would see a spot with the disorder insensitive profile and the
integral intensity $\propto e^{- f/l}$ (the exponential factor is
probability that the wave does not experience any scattering).

\subsection{Incoherent wave: the Boltzmann equation}
\label{Boltzmann}

Consider the initial density matrix of the form $\rho_{in}(y_{1},
y_{2})= \rho_{0}(y_{1}-y_{2})$, $y= y_{1}-y_{2}$, which corresponds to
a partially coherent spatially homogeneous state.  At $x>0$, the
density matrix $\rho (y_{1},y_{2};x)= \rho (y ;x)$ is found from
Eq.(\ref{6da}) with the two-particle propagator from Eq.(\ref{ad9}).
The Fourier transform,
\begin{equation}
\rho(y;x)= \int\limits_{-\infty }^{\infty }  d\varphi \, e^{i
\varphi {y\over{\lambdabar }}} \,n_{\varphi }(x ) \ ,
\label{xda}
\end{equation}
defines $ n_{\varphi }$ which has the meaning of the distribution
function with respect of the transverse momentum, $q= \varphi
{\hbar\over{\lambdabar}}$, and $\varphi $ is the angle between the
velocity of the particle and the $x$-axis.  In the paraxial situation,
the distribution is concentrated at small angles and the integration
in Eq.(\ref{xda}) can be taken in infinite limits.

From, Eqs.(\ref{6da}) and (\ref{ad9}) we easily get the density matrix 
of the
wave having traveled the distance $x$:
\begin{equation}
\rho (y;x)= \rho _{0}(y) \exp \left({- {x \mid y \mid\over{2 {\cal
L}^{2}}}} \right) \ .
\label{amh}
\end{equation}
As required by the current conservation, $\rho (0;x)$ does not depend
on $x$ whereas the non-diagonal elements of the density matrix $y\neq
0$ decay to zero, the faster, the more ``distance to the diagonal''
$|y|$. In other words, random magnetic field is very effective in
destroying a long range coherence, the longer the coherence, the
faster it decays.

Considering the limiting case of plane infinite incident wave
$n_{\varphi}(x=0)= \delta (\varphi )$ {\it i.e.}  $\rho_{in}(y)=1$,
Eq.(\ref{amh}) gives
\begin{equation}
n_{\varphi}(x)= {1\over \pi } {\Delta\over \varphi^{2} +
\Delta^{2}}
\;\; , \;\;  \Delta = x {\lambdabar \over 2 {\cal L}^{2}} 
\label{4da}
\end{equation}
The evolution is nonperturbative in the sense that the plane wave
looses its shape immediately at any $x\neq 0$ transforming into the
Lorentz distribution with the width of the distribution proportional
to $x$ and the strength of the field \cite{usual}.  In particular, it
means that the plane wave is not a good basis for the perturbation
theory.

More insight can be gained if the evolution of the density matrix is
mapped to a Boltzmann-type kinetic equation. 
For this, note that 
the density matrix in Eq.(\ref{amh}) satisfies the  equation,
\begin{equation}
v {\partial\over{\partial x}}\rho + \hat{I}\rho =0
\;\; , \;\;  \hat{I}= {v\over 2{\cal L}^{2}} |y| \ .
\label{yda}
\end{equation}
Written for the distribution function $n_{\varphi }$ introduced in Eq.(\ref{xda}),
the equation  acquires the familiar Boltzmann form, 
\begin{equation}
v {\partial n_{\varphi }\over{\partial x}} + \hat{I}n_{\varphi }=0 \ ,
\label{zda}
\end{equation}
where $\hat{I}$ is the collision integral {\it i.e.} operator $\hat{I}$
in the $\varphi $-representation.

One may 
present the collision integral in the standard form
\begin{equation}
\hat{I} n_{\varphi }=  \int d\phi  \;
 w(\phi)(n_{\varphi}- n_{\varphi + \phi })\; \; , 
\label{a629}
\end{equation}  
where $w(\phi )$ is the scattering rate for the process $\varphi
\rightarrow \varphi + \phi $.  From the condition that the operators
in Eq.(\ref{zda}) and Eq.(\ref{a629}) have same eigenvalues
corresponding to the common eigenfunctions, $n_{\varphi} \sim e^{i
{y_{0}\over \lambdabar } \varphi }$, $w$ must satisfy the requirement
that Ref.\cite{eigen}
\begin{equation}
\int\limits_{-\infty }^{\infty } d \phi \,w(\phi )(1- e^{-i
{y_{0}\over{\lambdabar }} \phi }) = {v\over 2 {\cal L}^{2}} |y_{0}| \ .
\label{bib}
\end{equation}
From here,
\begin{equation}
 w(\phi )= {2\over{\pi \tau_{0}}}  {1\over{\phi^{2}}}
\;\; , \;\;  
{1 \over \tau_{0}} = {v \lambdabar \over 4  {\cal L}^{2}}
\label{3da}
\end{equation}

Usually, one can split the collision integral into the in- and
out-scattering pieces. In the case of random magnetic field, the
scattering-out rate is ill-defined as $\int d \phi w(\phi )$ diverges
at small angles, and the split hardly makes sense. On the other hand,
the collision integral, as an operator acting on the distribution
function, is well defined and the transport is not singular.

Treating the random magnetic field in the Born approximation,
Aronov {\it et al.} 
\cite{AroMirWol94}  found the 
 the scattering rate $W(\phi )$ to be
\begin{equation}
W(\phi )= {1\over{2 \pi \tau_{0}}}  \cot^{2}{{\phi \over 2}} \ .
\label{7da}
\end{equation}

Because of divergence at small angles, one may doubt the validity of
the Born approximation. However, the small $\phi $ asymptotics of
$W(\phi )$ agrees with Eq.(\ref{3da}). This means that Eq.(\ref{7da})
is actually valid for arbitrary $\phi $ if used for constructing the
collision integral.  The other way around, the collision integral
Eq.(\ref{a629}) is expected to correctly describe scattering with
arbitrary scattering angles if $W(\phi )$ Eq.(\ref{7da}) is used
instead of paraxial $w(\phi )$ in Eq.(\ref{3da}).  This allows one to
generalize the paraxial kinetic equation, including large angle
scattering.

The kinetic equation for the distribution function  $n_{\varphi }$  reads
\begin{equation}
\bbox{v\cdot \nabla }n_{\varphi } + \hat{I} n_{\varphi }=0 \;\; , \;\;
 0< \varphi < 2\pi \ ,
\label{8da}
\end{equation}
where the collision integral
\[
 \hat{I} n_{\varphi }= {1\over  \tau_{0}} \int\limits_{0}^{2\pi }
 {d \phi \over 2 \pi  } \,  \cot^{2}{{\phi \over 2}} \, 
\left(n_{\varphi }- n_{\varphi + \phi }\right) \ . 
\]
As before, the spilt of the collision integral in the in- and out-
scattering parts leads to divergences and, therefore, has a very
limited sense.  At the same time, the collision operator is
well-defined: If $n_{\varphi}$ is presented as the sum,
\[
n_{\varphi}=  \sum\limits_{m  } n_{m} e^{im\varphi}\ ,
\]
over the eigenfunctions of the
collision operator $e^{im \varphi }$, $m= 0, \pm 1 ,  \ldots$, the
collision operator acts as  
\[
\hat{I}n_{\varphi}=  {1\over \tau_{0}} \sum\limits_{m \neq 0 }^{\infty }
(2 |m| -1) n_{m} e^{im\varphi} \; ,
\]
The parameter $\tau_{0}$ has the meaning of the relaxation time for
the first harmonics $m=\pm 1$ {\it i.e.} the transport relaxation
time.  

One concludes that transport of a charge in a random magnetic field
can be described by the Boltzmann equation, and is not anomalous
in spite of the fact that the total scattering rate is infinite.

\section{Random array of Aharonov-Bohm lines }
\label{AB-array}

This section deals with the model of a random gauge field where the
gauge field is created by an array of Aharonov-Bohm lines. It is
assumed that the lines in the array take random space positions and
the flux of a line $\Phi $ may be random.  The model is specified by
the averaged density of the lines $d_{_{\text{AB}}}$ and the
probability distribution $p(\Phi )$ for the magnetic flux $\Phi $ in a
line.  Our primarily goal is to average the paraxial two-particle
Green function Eq.(\ref{qda}) over the distribution of the lines.

The calculations turn out to be very similar to those in Section
\ref{random} so that we only outline them.  Repeating the arguments
from Section \ref{random} one comes to an expression similar to
Eq.(\ref{vda}):
\[
{\cal K}_{\text{av}}(y_{1},y_{2};x| y_{1}',y_{2};,x')=
{\cal K}_{\text{0}}(y_{1},y_{2};x| y_{1}',y_{2};,x')
\left< 
\exp \left( 2\pi i{\Phi^{(t)}(y_{1},y_{2};x| y_{1}',y_{2};,x')\over{\Phi_{0}}}
\right)\right> 
\]
where $\Phi^{(t)}(y_{1},y_{2};x| y_{1}',y_{2};,x')$ is the flux
through the oriented area bounded by the straight (directed) lines
connecting the initial and finite points (see
Fig.\ref{straight}). Given the configuration of the Aharonov-Bohm
array, the flux through the area is the sum over the lines piercing
the area.  The $k$'s line with the flux $\Phi_{k}$ contributes to the
total flux as $\sigma_{k}\Phi_{k}$ where $\sigma_{k}= +1$ or $-1$
depending on the orientation, positive or negative, of the area the
line is situated in.  Let $N_{+}$ ($N_{-}$) be the (random) number of
lines in the area with positive (negative) orientation; The variables
$\Phi_{k}$'s are independent in the model, and the averaging $\exp
\big(2\pi i {\Phi^{(t)}\over \Phi_{0}} \big)$ over the configurations
with fixed $N_{\pm}$ is simple:
\[
\left<   \exp \left(i {2\pi\over \Phi_{0}}
\sum\limits_{k=1}^{N_{+}+N_{-}} \sigma_{k}\Phi_{k}
\right)\right> = 
\left<   \exp \left(2\pi i  {\Phi \over \Phi_{0}}\right)\right>^{N_{+}}
\left<   \exp \left(-2\pi i  {\Phi \over \Phi_{0}}\right)\right>^{N_{-}}
\]
where $< \exp ({\pm {2\pi i\Phi \over \Phi_{0} }})>$ implies averaging
with the distribution function $p(\Phi)$.

The random numbers $N_{\pm}$ obey the Poisson distribution $P_{N}=
e^{\bar{N}} \bar{N}^{N}/N!$ with $\bar{N}$ either
$\overline{(N_{+}+N_{-})}= d_{_{\text{AB}}} {\cal A}_{\text{no}}$ or
$\overline{N_{+}-N_{-}}= d_{_{\text{AB}}} {\cal A}_{\text{o}}$, ${\cal
A}_{\text{no}}$ and ${\cal A}_{\text{o}}$ being non-oriented and
oriented area, respectively.

Finally,
\[
\left< 
\exp \left( 2\pi i{\Phi(y_{1},y_{2};x| y_{1}',y_{2};,x')\over{\Phi_{0}}}
\right)\right> = 
\exp \left(
- {{\cal A}_{\text{no}}\over{2\cal L}_{AB}^{2}}
+ i {2\pi \over \Phi_{0} } \tilde{B} {\cal A}_{\text{o}} 
\right) \ ,
\]
where  ${\cal A}_{\text{no}}$ is the non-oriented area Eq.(\ref{eea}),
${\cal A}_{\text{o}}$ is the oriented area,
\[
{\cal A}_{\text{o}}= - {1\over{2}}(x-x')(y_{1}+ y_{1}'- y_{2} - y_{2}')
\ , 
\]
\begin{equation}
{1\over{\cal L}_{AB}^{2}}=
2 d_{_{\text{AB}}} \left<1- \cos 2 \pi  {\Phi \over \Phi_{0}}\right>
\label{fea}
\end{equation}
and  the effective magnetic field 
\begin{equation}
\tilde{B}=  d_{_{\text{AB}}} {\Phi_{0}\over 2\pi } 
\left< \sin 2 \pi  {\Phi \over \Phi_{0}} \right> \ ;
\label{gea}
\end{equation}
in Eqs.(\ref{fea}) and (\ref{gea}), the averaging is performed with
the distribution function of the flux in the line $p(\Phi )$.

Collecting the results together, the two-particle Green function reads
\begin{eqnarray}
 K_{\text{av}}(y, Y; y', Y'; x )     & =&  \nonumber  \\
&&
\hspace{-20ex}
={1\over{2 \pi x \lambdabar}}
\exp \left[{i\over{x \lambdabar}}(Y-Y') (y-y')
-{1\over{8}} {x\over{{\cal L}_{_{\text{AB}}}^{2}}} 
\Big(\mid y \mid + \mid y'
\mid 
+ {(y+y')^{2}\over{\mid y \mid}+\mid y' \mid}\Big)
 - i {\pi \over \Phi_{0}} \,\tilde{B}\, x (y+y')
  \right] 
\label{jea}
\end{eqnarray}
where  
$y= y_{1}-y_{2}$, 
$y'= y_{1}'-y_{2}'$, 
$Y= {1\over{2}}(y_{1}+ y_{2})$ 
$Y'= {1\over{2}}(y_{1}'+ y_{2}')$ 
and $x \leftarrow (x-x')$.

If compared with the Green function derived in Sect.\ref{random}
Eq.(\ref{ad9}), the propagator Eq.(\ref{jea}) contains an additional
term proportional $\tilde{B}$, finite to the extent the distribution
$p(\Phi )$ is asymmetric.  As we will see later, $\tilde{B}$ creates
the Lorentz force and plays in dynamics the role of an effective
magnetic field.  A finite Lorentz force in an Aharonov-Bohm array is
not readily obvious: A classical Lorentz force is absent since
magnetic field is locally zero, whereas the quantum Aharonov-Bohm
cross-section Eq.(\ref{7hb}) is left-right and $\Phi \rightarrow -\Phi
$ symmetric and cannot explain $\tilde{B}$.  Obviously, $\tilde{B}$
and the associated force is directly related to the transverse
momentum transfer considered in Sect.\ref{AB} (see Eq.(\ref{ida})).

In the limit of dense array with small typical flux, $\Phi \rightarrow
0$, $d_{_{\text{AB}}}\rightarrow \infty $,  the effective field
$\tilde{B}$ reduces to 
 to the macroscopic mean-field
magnetic induction 
$B=d_{_{\text{AB}}}<\Phi >$.
In this limit, the Aharonov-Bohm array model is
equivalent the  $\delta -$correlated field model Eq.(\ref{vhb}) with
${\cal L}^{-2} 
\propto d_{_{\text{AB}}}<\Phi^{2}>$, 
in the external homogeneous magnetic field $\tilde{B}$.

In general, however, $\Phi$-periodic $\tilde{B}$ Eq.(\ref{gea}) is
very different from the mean-field expectations.  Depending on the
flux distribution function $p(\Phi )$, the magnetic induction and
$\tilde{B}$ may be in any relation. For instance, the mean-field
induction can be always compensated to zero by adding some properly
oriented lines of flux $\Phi_{N}= {N\over2 } \Phi_{0},\, N= 1,2,\ldots
$.  However, the added lines do not affect the effective magnetic
field $\tilde{B}$ seen by the particles (since $\sin
\left(2\pi{\Phi_{N}\over \Phi_{0}} \right)=0 $). In an Aharonov-Bohm
array the Lorentz force may be finite even when the macroscopic
magnetic induction is zero.

\subsection{Kinetic equation}\label{kin}

As before, in Sect.\ref{Boltzmann}, consider the spatially uniform
situation when the density matrix of the partially coherent wave at
$x=0$ is of the form $\rho_{in}(y_{1},y_{2})= \rho_{0}(y)$, $y=
y_{1}-y_{2}$. The density matrix $\rho (y;x)$ of the wave at distance
$x$ can be found from Eqs.(\ref{6da}) and Eq.(\ref{jea}),
\[
\rho (y;x)= \rho _{0}(y) \exp \left({- {x \mid y \mid\over{2 {\cal
L}_{AB}^{2}}}}
 - i {2\pi \over \Phi_{0}} \,\tilde{B}\, x y
\right) \ .
\]
Again, as in Sect.\ref{Boltzmann}, the density matrix obeys the
following kinetic equation (compare with Eq.(\ref{yda})):
\[
v {\partial\over{\partial x}}\rho + 
i {2\pi \over \Phi_{0} }\tilde{B} y \rho + \hat{I}_{_{AB}}\rho =0
\;\; , \;\;  \hat{I}_{_{AB}}= {v\over 2{\cal L}_{AB}^{2}} |y| .
\]
Performing the Fourier transform, one gets the equation
for the distribution function $n_{\varphi}$  in  Eq.(\ref{xda}):
\begin{equation}
v {\partial n_{\varphi }\over{\partial x}} + {e \tilde{B}\over mc }
{\partial n_{\varphi}\over{\partial \varphi  }}
+ \hat{I}_{_{AB}}n_{\varphi }=0 \ .
\label{mea}
\end{equation}
The equation has the Boltzmann form and $\tilde{B}$ enters kinetics as
a magnetic field.
The collision
integral $\hat{I}_{_{AB}}$ has the form of Eq.(\ref{a629}) with the
scattering rate
\[
W(\phi )= 
{v \lambdabar \over 2\pi {\cal L}^{2}_{_{AB}}}{1\over{\phi^{2}}}
\]
From here one can conclude that 
the relaxation
is governed by an incoherent
scattering by the flux lines: Indeed, $W$ is proportional to the
density of lines and the contribution of a line is given by the small
limit of the Aharonov-Bohm cross-section Eq.(\ref{7hb}). 
Since even most dangerous small angle scattering fits this simple
picture, it seems plausible that Eq.(\ref{mea}) can be generalized to
arbitrary scattering angle using Eq.(\ref{7hb}) as the probability
scattering.  Similar to Eq.(\ref{8da}), the kinetic equation reads
\[
\bbox{v\cdot \nabla }n_{\varphi } 
 - {e \tilde{B}\over mc } {\partial n_{\varphi}\over{\partial \varphi }} +
  \hat{I}_{_{AB}}n_{\varphi }=0 \;\; , \;\;
 0< \varphi < 2\pi \ ,
\]
where the collision integral
\[
 \hat{I}_{_{AB}}n_{\varphi }= 
{1\over 2 \tau_{_{AB}}} \int\limits_{0}^{2\pi }
 {d \phi \over 2 \pi  } \, {1 \over \sin^{2}{{\phi \over 2}}} \, 
\left(n_{\varphi }- n_{\varphi + \phi }\right) \ . 
\]
with
\begin{equation}
{1\over \tau_{_{AB}}} =  
{\hbar \over m }
d_{_{\text{AB}}} 
\left<1- \cos 2 \pi  {\Phi \over \Phi_{0}}\right> \ .
\label{qea}
\end{equation}
As before,  the collision integral 
is a regular linear operator regardless  its singular
scattering-out term.  Its action  is defined
by the following relation
\[
 \hat{I}_{_{AB}}e^{im \varphi }= {|m|\over \tau_{_{AB}}}\,\,e^{im \varphi }  
\]
From here, one sees that $\tau_{_{AB}}$ has the meaning of the
transport scattering time.

Applicability of the Boltzmann equation requires the mean free path $l
\sim v \tau_{_{AB}}$ to be large on the scale of the wave length
$\lambdabar$.
\begin{equation}
{\lambdabar \over l } \sim \lambdabar^{2} d_{_{\text{AB}}}
\left<\sin^{2}{\pi{\Phi\over\Phi_{0}}}\right> \ll 1  
\label{xea}
\end{equation}

If typically $\Phi/ \Phi_{0} \sim 1$, the density of the lines must
not be too high: $d_{_{\text{AB}}}\lambdabar^{2}\ll 1$. In case of
lines with small $\Phi $, the condition is milder:
$\Phi^{2}d_{_{\text{AB}}}\lambdabar^{2}\ll 1$.

As an illustration, the Drude conductivity tensor can be readily derived
from the Boltzmann equation Eq.(\ref{qea}): 
\begin{eqnarray}
\sigma_{xx}     & =  &    
{1\over{2}} \,e^{2}\,N_{0}\, v^{2} \tau_{_{AB}} {1\over 1 +
(\Omega_{c}\tau_{_{AB}})^{2}}          \label{9ib}\\   
\sigma_{xy}&= &(\Omega_{c}\tau_{_{AB}}) \sigma_{xx}
\label{9ib2}   
\end{eqnarray}
where $N_{0}$ is the density of states and $\Omega_{c}= {e
\tilde{B}\over mc }$ plays the role of the Larmor frequency. One sees
that the Hall angle $\theta_{H}$, $\tan\theta_{H}= \sigma_{xy}/
\sigma_{xx}$,
\begin{equation}
\tan\theta_{H} = {\left< \sin 2\pi{\Phi \over \Phi_{0}} \right>  \over
\left< 1 - \cos 2\pi{\Phi \over \Phi_{0}} \right> }
\ ,
\label{sea}
\end{equation}
does not depend on the density of lines and parameters of the
system. If the Aharonov-Bohm lines in array have same flux $\Phi $,
the Hall angle is simply
\begin{equation}
\tan\theta_{H} = \cot \left(\pi  {\Phi \over \Phi_{0}} \right)  \ .
\label{rea}
\end{equation}
Counter intuition, 
the Hall effect is strongest at $\Phi \rightarrow 0$.

\subsection{Landau quantization}\label{Landau}

The Hall angle Eq.(\ref{rea}) in an Aharonov-Bohm array may be large
which means that the particle orbit tends to be a circle, similar to
Larmor orbits in an external magnetic field. The periodic classical
motion is expected to be quantized (the Landau quantization). To check
this possibility, one should calculate the density of states of
a particle moving in an Aharonov-Bohm array.  In the quasiclassical
region, when the energy of the particle $E\gg \hbar \Omega_{c}$, the
problem can be generally solved by the method used in
\cite{AroAltMir95-1}.  In the present paper, only the most promising
case of small flux array is considered.  Not to repeat the calculation
in \cite{AroAltMir95-1}, qualitative arguments which lead to the same
result, are presented.

In the linear in $\Phi $ approximation, when scattering $\propto
\Phi^{2}$ can be neglected, the transverse momentum Eq.(\ref{0hb}) is
transfered to the particle, and it undergoes periodic motion with the
angular frequency $\Omega_{c}= {e \tilde{B}\over mc }$ along the
Larmor circle with the radius $ R_{L}= v/\Omega_{c}$.  In the small
$\Phi $-limit, the distinction between $\tilde{B}$ and magnetic
induction is immaterial, and the Bohr-Sommerfeld quantization
condition can be formulated as the requirement of the flux
quantization,
\begin{equation}
\Phi^{\text{orbit}}(E_{N}) = \Phi_{0}(N + {1\over{2}}) \ ,
\label{yea}
\end{equation}
where $\Phi^{\text{orbit}}(E) $ is the total flux encircled by an
orbit of the particle with energy $E$.  Due to the randomness in the
flux line positions, the number of the encircled lines fluctuates, and
so does the total flux.  The total flux $\Phi^{\text{orbit}}=
\langle\Phi^{\text{orbit}}\rangle+ \delta \Phi$ is a random Gaussian
variable which fluctuates around the average $\langle
\Phi^{\text{orbit}} \rangle =\tilde{B}S(E)$, $S(E)= \pi R_{L}^{2}$
being the Larmor circle area. For the flux lines with uncorrelated
positions, the total flux fluctuates with the variance
\begin{equation}
\langle\big(\delta \Phi
\big)^{2}\rangle= d_{_{\text{AB}}}S(E)\langle\Phi^{2}\rangle \ . 
\label{1ea}
\end{equation}
Driven by the flux enclosed by the orbit, the energy of the level
fluctuates, $E_{N}= \langle E_{N}\rangle + \delta E$.  Neglecting the
flux fluctuations in Eq.(\ref{yea}), $\Phi^{\text{orbit}} \rightarrow
\langle\Phi^{\text{orbit}}\rangle $, one gets the average energy of
the N-th level $\langle E_{N}\rangle= \hbar \Omega_{c}(N+
{1\over{2}})$.
To preserve the quantization Eq.(\ref{yea}), 
the level energy acquires a shift $\delta E$ under 
the flux variations 
$\delta \Phi $:
From the condition
\[
\delta \Phi + \delta E_{N} {\partial \Phi  \over{\partial E
}}|_{E_{N}}=0
\; ,
\]
one gets the energy shift caused by the change of the flux $\delta \Phi $:
\begin{equation}
\delta E_{N}= - \hbar \Omega_{c} {\delta  \Phi \over \Phi_{0} } \ .  
\label{fjb}
\end{equation}
Combining Eq.(\ref{fjb}) and Eq.(\ref{1ea}), we see that the Landau
levels acquire the Gaussian distribution form with variance (the width
of the level)
\begin{equation}
\left<\big(\delta E_{N} \big)^{2}\right>= {1\over \pi }\, 
{\hbar \over \tau_{_{AB}}}\, E  
\label{zea}
\end{equation}
Physics here is similar to the inhomogeneous broadening: The Larmor
circle sees different realizations of the random relief in different
places of the ``sample'', and the energy levels adjust their positions
to the local conditions.

As already noticed in \cite{AroAltMir95-1},
it is rather unusual that importance of disorder (measured by the Landau
level broadening Eq.(\ref{zea})) increases with the energy $E$.
The physical reason for this  is that the larger is proportional
to $E$ the area under the Larmor circle, the bigger are the absolute
fluctuations of the number of the Aharonov-Bohm lines encircled by the
orbit and the flux fluctuation $\delta \Phi $.

The density of states is given by the sum over the discrete
levels. The position of the level is the Gaussian of the width in
Eq.(\ref{zea}) and centered at $E_{N}= \hbar \Omega_{c}(N+
{1\over{2}})$.  The overlapping Landau levels create oscillating
density of states. Similar to Ref.\cite{AroAltMir95-1}, one applies
the Poisson summation formula and finds the first harmonics of the
oscillations:
\[
\rho^{osc}(E)= - {m\over \pi \hbar^{2} }\exp (-\gamma ) \cos
\left(2\pi  {E\over \hbar \Omega_{c} } \right) \;\; ,
\]
where the damping of the oscillations is controlled by 
\[
  \gamma =
{2\pi \over \Omega_{c} \tau_{_{AB}} }\, {E\over \hbar \Omega_{c} } \ .
\]
Therefore, the Shubnikov-De-Haas oscillations in the Aharonov-Bohm
array are strongly suppressed: even when $\Omega_{c} \tau_{_{AB}} =
\tan \theta_{H} \gtrsim 1$ and the Larmor circling is well pronounced,
quasiclassical quantization at high Landau levels $E \gg \hbar
\Omega_{c}$ may not be seen because of the large damping $\gamma $.

The damping parameter $\gamma $ can be also presented as 
\begin{equation}
\gamma = {\pi \over 2 } \,
{1\over\lambdabar^{2} d_{_{\text{AB}}}}\, 
{\langle\Phi^{2}\rangle\over \langle\Phi \rangle^{2}} 
\,\,\,\geq \,\,\,{\pi \over 2 } {1\over
\lambdabar^{2} d_{_{\text{AB}}}}
\label{3ea}
\end{equation}
where $\lambdabar $ ($= \hbar /\sqrt{2m E}$) is the wave length
corresponding to the energy $E$; the equality sign realizes in the
case when the lines have equal flux. One concludes from Eq.(\ref{3ea})
that as a necessary condition, the Larmor motion may be quantized only
if the wave length exceed the distance between Aharonov-Bohm lines. As
one could expect, this requirement tends to be complementary to the
condition of applicability of the Boltzmann equation in
Eq.(\ref{xea}).

\section{Conclusions}
\label{concl}

The main goal of the paper has been to understand specific features of
the transport properties of a quantum charge subject to a random
magnetic field, namely, the features related to the long range
correlations of the gauge potential and the anomalous forward
scattering.  The non-perturbative method used in the paper to handle
the divergence is based on the paraxial approximation to the
stationary Schr\"odinger equation (Section \ref{parax}).

The paraxial theory of magnetic scattering is presented in Section
\ref{magn}. To show usage of the theory, it is applied to the
Aharonov-Bohm line problem in Section \ref{AB}.  Being calculationally
simple, the paraxial approximation proves to be rather efficient.  The
paraxial solution reproduces the small angle asymptotics of the 
Aharonov-Bohm exact solution for the plane incident wave. 
The wave packet solution (see Eq.(\ref{6hb})) allows one to resolve an
old controversy
discussed in the Introduction
concerning the transverse force exerted by the
Aharonov-Bohm line.  One sees that the angular distribution in the
outgoing wave is indeed left-right {\it asymmetric}, so that there is
a finite momentum transfer in the transverse direction. However, the
asymmetry is concentrated within the angular width of the incident
wave and it cannot be described in terms of the differential
cross-section. For an arbitrary incident wave, the transverse momentum
transfered to the charge can be found by Eq.(\ref{0hb}).  By
comparison with the exact solution, the validity of this formula has
been recently confirmed by Berry \cite{Ber99}.

The main result of the paper is the paraxial two-particle Green's
function averaged with respect to the random magnetic field.  In the
paraxial theory, a stationary 2D problem becomes equivalent to a
non-stationary 1D one, and one can use the standard Feynman
representation for the propagators.  It turns out that the
corresponding path integral can be evaluated exactly.  The expressions
for the Green's function is given by Eqs.(\ref{ad9}) and
Eq.(\ref{jea}) for the Gaussian random magnetic field and the
Aharonov-Bohm array model, respectively.

The paraxial two-particle Green's function solves the quantum problem
of the near forward multiple scattering by random gauge potential: for
a given incident wave, one is able to find correlators $\langle \psi
(\bbox{r}_{1}) \psi^{*}(\bbox{r}_{2})\rangle $, where averaging is
performed with respect to the random field.  To draw physical
conclusions, two cases are analyzed: (i) (de)focusing of a coherent
converging wave in the random magnetic field environment, and (ii)
propagation of spatially homogeneous partially coherent beam.

The non-local character of interaction with magnetic field is clearly
seen from the analysis in Section \ref{focus} of defocusing of a
converging beam. The lost of coherence measured by defocusing is
controlled by the size of the entire area ``occupied `` by the system
that is the 
region where the wave function is finite: Indeed,
it follows from Eq.(\ref{wr}) that the wave cannot be focussed if the
random flux threading the area 
(aperture size)$\times$(focus length) 
is of order of the flux quantum $\Phi_{0}$.  Same conclusion follows
from Eq.(\ref{4da}): having traveled a distance $x$, a perfect plane
wave becomes a mixture of waves with the transverse momenta $\Delta
p_{y} \sim \hbar x / {\cal L}^{2}$ where the spatial coherence
survives only within the region 
$\Delta y \sim {\cal L}^{2}/x$. Again, supporting the qualitative
arguments presented in the Introduction, the coherence exists only
within the area $\Delta x \times \Delta y \sim {\cal L}^{2}$; the area
is defined by the condition that the
random flux is typically
not bigger than the flux quantum $\Phi_{0}$.

On the other hand, the evolution in the momentum space
is rather ordinary: It is described by the usual Boltzmann equation 
derived in Sections \ref{Boltzmann} and \ref{kin}.  The only uncommon
feature of the Boltzmann equation is that the collision integral being
a well-defined operator nevertheless cannot be split in the scattering
-in and -out {\it regular} parts: An attempt of the split produces two
singular pieces, the two infinities canceling when combined. In the
diagrammatic language, this can be rephrased as the cancellation of
the divergences in the self energy and the vertex correction as
observed in Ref.\cite{KimFurWen94}.

The condition of  applicability of the paraxial approximation can be
derived from Eq.(\ref{oca}):  At the distance $x$, the typical
transverse size $w$ is of order $min \,[w_{0}, {\cal L}^{2}/x]$ where 
$w_{0}$ is the characteristic length in the initial distribution. 
For large enough $x$,  $w \sim {\cal L}^{2}/x$. Substituting this
value into Eq.(\ref{oca}), one gets
\[
x < x_{max}\;\; , \;\;  x_{max}\equiv {\cal L} \left({{\cal L}\over
\lambdabar } \right)^{{3\over 5}} \; .
\]
One sees that the theory is applicable in
the non-perturbative region $x \gg {\cal L}$ in the paraxial limit
$\lambdabar \ll {\cal L} $.

In the paraxial picture, the particle always moves in (almost) same
direction. Therefore, any effect related to the Anderson localization
is beyond the paraxial approximation.  Although the Boltzmann equation
allows for large angle scattering events, it is, of course, unable to
describe the quantum  localization either.  Localization  in
random magnetic field remains a controversial issue, see
Refs.\cite{AroMirWol94} and \cite{TarEfe00}.

Two models of the random field has been considered in the paper: the
Gaussian random magnetic field with zero average and the array of
Aharonov-Bohm magnetic flux lines with arbitrary distribution of the
line fluxes.  Comparing the Green's function in Eqs.(\ref{ad9}), and
(\ref{jea}) (with $\tilde{B} \rightarrow 0$), one sees that the models
are paraxially equivalent.  (For the Gaussian model, Eq.(\ref{ad9})
has been derived assuming that the field is zero on average, $\langle
b \rangle=0 $. In a more general case, the Gaussian model with a
finite magnetic induction $B= \langle b \rangle $ has the Green's
function of the form in Eq.(\ref{jea}) with $\tilde{B}$ substituted
for $B$).

The origin of the effective magnetic field in the Aharonov-Bohm array
can be traced back to the left-right asymmetry in the scattering by an
isolated Aharonov-Bohm line (see Section \ref{AB}). One sees that the
transverse momenta $\Delta p_{\perp}$ Eq.(\ref{0hb}) gained as a
result of collisions with Aharonov-Bohm lines, add together giving
rise to the Lorentz force and the effective magnetic field $\tilde{B}$
Eq.(\ref{gea}).  Since any integer flux can be gauged out, $\Delta
p_{\perp}$ and $\tilde{B}$ are periodic functions of the fluxes.  Note
that the magnetic induction $B$, which by definition equals to
$\langle \Phi \rangle $, and the effective field $\tilde{B}$
Eq.(\ref{gea}) are same quantities only if lines flux is small.
Generally, they may be in any relation.
In particular, one may have finite
$\tilde{B}\neq 0$ even if the magnetic induction $B=0$: A complex of 3
lines with the fluxes: $(+ {\Phi_{0}\over 4}, + {\Phi_{0}\over 4}, -
{\Phi_{0}\over 2}$) gives an example.

In Section \ref{kin}, the kinetic equation for a charge in an
Aharonov-Bohm line array is solved to find the Drude conductivity
tensor Eq.(\ref{9ib}-\ref{9ib2}) and the Hall angle Eq.(\ref{sea}).
These results for the array of lines are in agreement with
Ref.\cite{DesOuvTex97} where the Hall effect due to a single line has
been studied.  As discussed in \cite{NieHed95}, the Aharonov-Bohm
periodicity, $\tilde{B}_{\Phi + \Phi_{0}} = \tilde{B}_{\Phi}$ combined
with the time reversal symmetry, $\tilde{B}_{\Phi}=
-\tilde{B}_{-\Phi}$, requires that the half-integer flux lines,
$\tilde{\Phi}= {N\over 2}$ do not generate any Lorentz force.  The
Abrikosov vortex carries the flux ${1\over 2} \cdot {hc\over e}$,
therefore, does not exert the transverse force (of course, only in the
limit when the particle wave length much exceeds the vortex size). As
noticed before, this is in a qualitative agreement with the
experimental observation of the reduced Hall effect exhibited by 2D
electrons in the magnetic field of the Abrikosov vortices
\cite{GeiDubKha90,GeiBenGri92,GeiBenGri94}.

In an array of Aharonov-Bohm lines with small fluxes $\Phi \ll
\Phi_{0}$, the transverse momentum transfer $\propto \Phi $ (the
``Lorentz'' force exerted by $\tilde{B}$) is more efficient than
relaxation of momentum the rate of which is ${1\over
\tau_{\text{AB}}}\propto \Phi^{2}$.  Therefore, the ``Lorentz'' force
significantly bends the particle trajectory leading to a large Hall
angle Eq.(\ref{rea}) and, probably, to Landau quantization. As shown
in Section \ref{Landau}, these expectation are almost never met. Even
if the Hall angle is large and the Larmor circling is well pronounced,
the Landau levels are very broad. The inhomogeneous in its nature
broadening is due to the fluctuations in the total flux threading the
Larmor orbit.
  
Summarizing, the analysis of the forward scattering anomalous
scattering in a random magnetic field (Gaussian and the Aharonov-Bohm
array) has been presented using the paraxial approximation to the
stationary Schr\"odinger equation. The gauge-invariant two-particle
Green's function has been found by exact evaluation of the
corresponding Feynman integral.  The propagation of coherent
(defocusing) and incoherent (Boltzmann equation) waves has been
analyzed, as well as Landau quantization in the Aharonov-Bohm array.

\section{Acknowledgements}
I am grateful to A. Mirlin and P. W\"olfle for discussions, and to
M. Ozana for the help with preparation of the paper.  This work
was supported by SFB 195 der Deutschen Forschungsgemeinschaft and
partly Swedish Natural Science Research Council.

\appendix

\section{The transverse force} \label{trans}

The transverse force that is the transfered momentum in the direction
perpendicular to the velocity, can be easily found in the paraxial
theory.  The expectation value of the transverse momentum in the
outgoing wave is
\begin{equation}
\langle \hat{p}_{y}\rangle_{out}= 
{\hbar\over{i}}\int\limits_{-\infty}^{\infty}dy\, \psi^{*}_{\text{out}}(x,y) {\partial\over{\partial y}}\psi_{\text{out}}(x,y)
\label{gda}
\end{equation}
To calculate this integral, one exploits the fact that  the free
propagation at $x>0$ conserves the momentum, and, therefore, the
average in Eq.(\ref{gda}) does not depend on $x>0$. The goal now is to
present it as an integral at $x=0$ where the $\psi_{\text{out}}$
in Eq.(\ref{vca}) is simplest possible; it cannot be done directly
because of the eikonal discontinuity in $\psi_{\text{out}}$ at $y=0$

First, note that $\psi_{\text{out}}(x,y)$ in Eq.(\ref{rca}), is a
well-behaved function of $y$ at any finite $x>0$.  Thus, the derivate
${\partial\over{\partial y}}$ in Eq.(\ref{gda}) can be safely taken as
the limit: ${\partial\over{\partial y}}f(y)= {1\over{2\eta
}}\left(f(y+ \eta)- f(y- \eta ) \right), \, \eta \rightarrow 0 $, and
Eq.(\ref{gda}) transforms to
\begin{equation}
\langle \hat{p}_{y}\rangle_{out}= 
{1\over{4\eta }} \left(
P_{out}(\eta )-
P_{out}(-\eta ) 
 \right) \;\; , \;\;  \eta \rightarrow 0
\label{hda}
\end{equation}
where
\[
P_{out}(\eta )=  \int\limits_{-\infty }^{\infty }dy\,
\psi_{\text{in}}^{*}(x,y_{-})\, \psi_{\text{in}}(x,y_{+})\;\; , \;\;
y_{\pm}= y \pm {1\over{2}}\eta  \ .
\]
Substituting everywhere the subscript $out \rightarrow in $, one gets
the transverse momentum in the incoming wave $\langle
\hat{p}_{y}\rangle_{in}$ expressed via the corresponding $P_{in}(\eta
)$.

Identically, $P_{out}(\eta )= \langle e^{{\hbar\over{i}}\eta
\hat{p}_{y}}\rangle $. Since $P_{out}$ is the expectation value of the
conserving variable $e^{i\eta \hat{p}_{y}}$, its value does not depend
on $x$.  Choose the point $x=+0$ where $\psi_{\text{out}}$ is given by
Eq.(\ref{vca}) to evaluate $P_{out}$.  After simple calculations,
\[
P_{out}(\eta )= 
P_{in}(\eta )+ |\eta | \left(e^{2\pi \hat{\eta }i {\Phi
\over{\Phi_{0}}}}-1\right) \overline{\,|\psi_{\text{in}}|^{2}_{\eta }} \ ,
\]
where
\[
 \overline{\,|\psi_{\text{in}}|^{2}_{\eta }}= {1\over{\eta }}
\int\limits_{-{\eta \over{2}}}^{{\eta \over{2}}} 
\psi_{\text{in}}^{*}(y_{-})\, \psi_{\text{in}}(y_{+})
\]
Now, we see that the limit in Eq.(\ref{hda}) is well defined, and we
get for the transfered momentum $\Delta p_{y}= \langle \hat{p_{y}}
\rangle_{out} - \langle \hat{p_{y}} \rangle_{in} $
\begin{equation}
\Delta p_{y} = \hbar  |\psi_{\text{in}}(0)|^{2} \sin 2\pi{\Phi\over{\Phi_{0}}}
\label{ida}
\end{equation}
This the final result for the transfered momentum.


\begin{references}
\bibitem{footnote99} 
Note that $G^{R}_{0}$
as a function of $y$  is concentrated at $|y|\sim \sqrt{\lambdabar x}$.
In agreement with the estimate in Eq.(\ref{nca}),
the transverse structure of the beam 
is changed at
$x\sim {w^{2}\over{\lambdabar }}$, $w$ being the length
associated with the structure.
\bibitem{footnote2}
From Eqs.(\ref{9ba}) combined with (\ref{0ba}) and (\ref{dca}), one
gets Eq.(\ref{eca}) with the r.h.s. multiplied by $\exp[ -i {e\over
\hbar c} \int_{(x_{out}, y_{*})}^{(x_{in},y_{*})} dx\; A_{x}]$.  This
immaterial global phase factor is omitted in Eq.(\ref{eca}).
\bibitem{ffoot}. As discussed earlier in Sect.\ref{magn}, the overall
phase of $\psi_{out}$ does not have physicall meaning.
Without changing any physics, $\psi_{out}$ is multiplied by by 
$\exp[i \pi \tilde{\Phi }]$ for convenience.
\bibitem{footnote11} The simplest way to check them is to use
Eq.(\ref{5ca1})).
\bibitem{footnote6ca}
Differentiating the l.h.s. of Eq.(\ref{5ca}) with respect to $x$, one
observes that the derivative vanishes by virtue of
Eq.(\ref{6ba}). This means that the l.h.s. is $x$ independent at $x>
x'$ and can be calculated in the limit $x \rightarrow x'$.  In this
limit $G^{R}(\bbox{r},\bbox{r}') \theta (x-x') \rightarrow \delta
(y-y')$ and Eq.(\ref{5ca}) follows immediately.
\bibitem{usual}
Usually, when the scattering time $\tau $ is finite, the initial
distribution $n_{0}$ decays as $n_{0}\exp[- t/ \tau ]$. E.g. in the
simplest case of the isotropic scattering, distribution function has
the structure $n(x)= n_{0}\exp[- t/ \tau ] + \left(1- \exp[- t/ \tau
]<n_{0}>\right) $. Here, the distribution smoothly transforms to the
initial at $t \rightarrow 0$, and for $t \ll \tau $ scattering can be
described by the perturbation theory.
\bibitem{eigen} One sees from Eq.(\ref{yda}) that the eigenfunctions,
$|y_{0}\rangle$, $\hat{I}|y_{0}\rangle = y_{0} |y_{0}\rangle $ are
$\delta $-functions in the $y-$representation, $\langle y|
y_{0}\rangle = \delta (y-y_{0})$, and therefore, in the $\varphi
$-representation, $\langle \varphi | y_{0}\rangle =e^{i {y_{0} \varphi
\over \lambdabar}}$.  Then, $\hat{I} e^{i {y_{0} \varphi \over
\lambdabar}} = (v/2 {\cal L}^{2})y_{0}e^{i {y_{0} \varphi \over
\lambdabar}}$, and one comes to Eq.(\ref{bib}).
\end{references}
\end{document}